\documentclass[tikz,nofootinbib,superscriptaddress,eqsecnum,tightenlines,11pt]{revtex4}

\usepackage{hyperref}
\usepackage{graphicx}
\usepackage{amsmath,amssymb,amsfonts,amsthm,stmaryrd,mathtools}
\usepackage{multirow,bigdelim}
\usepackage{dsfont}
\usepackage{color}
\usepackage{tikz}
\usepackage{tikz-3dplot}
\usetikzlibrary{calc}
\usetikzlibrary{shapes.geometric,patterns} 
\usepackage{blkarray}
\usepackage[normalem]{ulem}

\makeatletter
\newcounter{subsubsubsection}[subsubsection]
\renewcommand\thesubsubsubsection{\thesubsubsection .\@alph\c@subsubsubsection}
\newcommand\subsubsubsection{\@startsection{subsubsubsection}{4}{\z@}%
                                     {-3.25ex\@plus -1ex \@minus -.2ex}%
                                     {1.5ex \@plus .2ex}%
                                     {\centering\normalfont\small\textit}}
\newcommand*\l@subsubsubsection{\@dottedtocline{3}{10.0em}{4.1em}}
\newcommand*{\subsubsubsectionmark}[1]{}
\makeatother
\newcommand\scalemath[2]{\scalebox{#1}{\mbox{\ensuremath{\displaystyle #2}}}}

\def\be{\begin{equation}}
\def\ee{\end{equation}}
\def\ba{\begin{eqnarray}}
\def\ea{\end{eqnarray}}
\def\bas{\begin{subequations}\begin{eqnarray}}
\def\eas{\end{eqnarray}\end{subequations}}

\def\q{\quad}

\def\nn{\nonumber}
\def\q{\qquad}

\def\i{\mathrm{i}}

\begin{document}

\title{ Holographic description of  boundary gravitons in (3+1) dimensions}

\author{Seth K. Asante}
\email{sasanteATperimeterinstitute.ca}
\affiliation{Perimeter Institute for Theoretical Physics, 31 Caroline Street North, Waterloo, ON, N2L 2Y5, Canada}
\affiliation{Department of Physics and Astronomy, University of Waterloo, 200 University Avenue West, Waterloo, ON, N2L 3G1, Canada}

\author{Bianca Dittrich}
\email{bdittrichATperimeterinstitute.ca}
\affiliation{Perimeter Institute for Theoretical Physics, 31 Caroline Street North, Waterloo, ON, N2L 2Y5, Canada}
\affiliation{ Institute for Mathematics, Astrophysics and Particle Physics, Radboud University, Heyendaalseweg 135, 6525 AJ Nijmegen, The Netherlands}

\author{Hal M. Haggard}
\email{haggardATbard.edu}
\affiliation{Physics Program, Bard College, 30 Campus Road, Annandale-On-Hudson, NY 12504, USA}
\affiliation{Perimeter Institute for Theoretical Physics, 31 Caroline Street North, Waterloo, ON, N2L 2Y5, Canada}

\begin{abstract}
Gravity is uniquely situated in between classical topological field theories and standard local field theories. This can be seen in the the quasi-local nature of gravitational observables, but is nowhere more apparent than in gravity's holographic formulation. Holography holds promise for simplifying computations in quantum gravity. While holographic descriptions of three-dimensional spacetimes and of spacetimes with a negative cosmological constant are well-developed, a complete boundary description of zero curvature, four-dimensional spacetime is not currently available. Building on previous work in three-dimensions, we provide a new route to four-dimensional holography and its boundary gravitons. Using Regge calculus linearized around a flat Euclidean background with the topology of a solid hyper-torus, we obtain the effective action for a dual boundary theory which describes the dynamics of the boundary gravitons. Remarkably, in the continuum limit and at large radii this boundary theory is local and closely analogous to the corresponding result in three-dimensions.  The boundary effective action has a degenerate kinetic term that leads to singularities in the one-loop partition function that are independent of the discretization. 
These results establish a rich boundary dynamics for four-dimensional flat holography. 
\end{abstract}

\maketitle


\section{Introduction}

The role of boundaries has become more and more important for various approaches to quantum gravity. Holographic dualities, e.g. the AdS/CFT framework, suggest that a theory of quantum gravity can be dually described by a  field theory defined on an asymptotic boundary.  On the other hand, recently  a lot of  attention has been focussed on boundary degrees of freedom, which might emerge through the breaking of diffeomorphism, or other gauge symmetries, by a boundary, e.g. \cite{Gervais:1976ec, BarnichBr, StromingerL,  DonnellyFreidel, Geiller:2017whh, Gomes:2018dxs}.   In particular, such boundary degrees of freedom  are thought to play a role in explaining black hole entropy, for instance  in \cite{Carlip1,Bala, Hawking,Carlip17,Barnich2}.

In three-dimensional quantum gravity the two themes,  holographic duality and  boundary degrees of freedom, merge together in an interesting way.   Carlip has worked out  a dual holographic boundary theory that arises from the breaking  of (normal) diffeomorphisms by the presence of the asymptotic boundary in 3D AdS gravity \cite{CarlipAdSHolo}.  In this paper we consider gravity without a cosmological constant. In this case one can also obtain dual boundary theories, not only at the asymptotic boundary \cite{Gombi,BarnichOneLoop}, but also for finite boundaries \cite{BonzomDittrich,qlh1,qlh2,qlh3,qlhA}.

We briefly review some developments that motivate the current work:  Barnich et al computed the one-loop partition function for 3D gravity  for a solid torus, in the limit of infinite radius \cite{BarnichOneLoop}. The result
\ba\label{3DRes}
{\cal Z}_1(\beta,\gamma) \,=\, e^{\frac{\beta}{8G}} \, \prod_{k=2}^\infty \frac{1}{|1-e^{\i\gamma k }|^2}
\ea
 depends on the moduli parameters $\beta$ and $\gamma$, with $\beta$ specifying the length of the axis of the solid torus, and $\gamma$ its Dehn twist.  The action $S=-\beta/8G$, with $G$ the 3D Newton's constant, is that of the flat solution, which arises solely from the boundary term.\footnote{This boundary term is $1/2$ of the usual Gibbons-Hawking-York-boundary term. This choice is justified for an asymptotically flat boundary in \cite{Detourney}.}  The second factor is a one-loop correction. It indicates that there are field modes present, despite the fact that 3D gravity is topological, in the sense of having no propagating degrees of freedom. The  explanation for this apparent paradox is that these field modes describe boundary degrees of freedom. In fact, this partition function (\ref{3DRes}) reproduces the 3D vacuum character of the $\text{BMS}_3$ group \cite{Oblak1,Oblak2,Oblak3}, which is an infinite-dimensional group describing the asymptotic symmetries of 3D asymptotically flat space.\footnote{An analogous result for the AdS case has been derived in \cite{Jensen}.}

How, then, does one identify the boundary degrees of freedom and the action that governs their dynamics?  The arguments of Carlip in \cite{CarlipAdSHolo} suggest that the geodesic distance, defined from a point on the boundary  to some central point,\footnote{The precise definition of this central point is inconsequential because its displacement will amount to a gauge transformation for the boundary field. In the case of the solid torus the central point is replaced by a central axis.}  could provide a suitable boundary field.  Indeed, these variables would describe how the boundary is embedded into the 3D (flat) solution. In particular, they would respond to deformations of the boundary normal to itself, that is, to boundary normal diffeomorphisms. It has become customary to refer to these degrees of freedom as boundary gravitons. 

 The geodesic distance can be understood as a functional of the metric, and hence we might ask for the effective dynamics, as induced by the 3D gravity action for this functional. 
 
 Regge calculus \cite{Regge}, a discretization of gravity in which the basic variables are the lengths of the edges of a triangulation of spacetime, turns out to be a very convenient framework, in which just such an effective dynamics can be computed. To this end one starts with a finite boundary and allows for the boundary metric to fluctuate, so that one can describe the boundary gravitons.  The discretization is also used as a regulator for the path integral, and allows a straightforward evaluation of the one-loop determinant, even if one considers regions with boundary \cite{BonzomDittrich}.

One reason why Regge calculus turns out to be so convenient for this task is that its 3D one-loop partition function is bulk triangulation invariant. Hence you can work with an arbitrarily coarse bulk triangulation. In particular, one can choose a triangulation with an edge connecting each boundary vertex to some central bulk vertex (or to vertices on some central axis if one considers a solid torus). Since the solutions to 3D gravity are flat one can identify these edges with geodesics.  To obtain an effective theory for the lengths of these edges one just needs to integrate out the lengths of all other bulk edges.

If the topology of the bulk spacetime is that of a ball there is even a triangulation in which all bulk edges are radial and go  from the boundary to some central point. Hence, the Regge action itself serves as an effective action for the geodesic length variables, without the need to integrate out any variables.  
The Regge action is local, and thus one obtains a local boundary field theory, whose partition function agrees with that of gravity.  Note that for other bulk topologies one might need to integrate out some set of edges and there is a priori no guarantee of finding a local boundary theory. Indeed, for the solid torus topology locality holds only in the large radius limit \cite{BonzomDittrich}.

We briefly review a few results from \cite{BonzomDittrich}. In the large radius limit, there exists a boundary field theory description, whose (linearized) action is characterized by a degenerate kinetic term (described by a quadratic form that can be obtained from the trace-reversed extrinsic curvature tensor)  and a coupling to the Ricci scalar of the boundary metric.\footnote{Similar boundary theories with degenerate kinetic terms were obtained in \cite{BarnichGonzales} and  \cite{CarlipMink}. These derivations were based on very different arguments from the ones used in \cite{BonzomDittrich}:  \cite{BarnichGonzales} relied  on  BMS symmetry  and \cite{CarlipMink} considered a null boundary with asymptotically  flat boundary conditions.}
 This boundary field theory explains the structure of the one-loop correction in (\ref{3DRes}).  Indeed, the partition function (\ref{3DRes}) has been reproduced in \cite{BonzomDittrich} using Regge calculus, and shown to be valid also for finite boundaries. This boundary field theory provides a reformulation of 3D gravity (without cosmological constant) as a theory describing  how  a (boundary) surface is embedded into 3D flat space. 

One might assume that this description, which makes use of the fact that the solutions of 3D gravity are flat, holds only at the perturbative level. 
The works \cite{qlh1,qlh2,qlh3,qlhA} show, however, that it also holds at the fully non-perturbative level. Here, one uses the Ponzano-Regge partition function for 3D gravity \cite{PR}, which provides a non-perturbative model for 3D quantum gravity. This model allows, in particular, the evaluation of the partition function for metric boundary conditions. More precisely, using techniques developed in loop quantum gravity, the boundary conditions are encoded in boundary wave functions. These wave functions can be chosen to exhibit the full range of possibilities from a deeply quantum to a completely semiclassical boundary.  Depending on this choice of wave function one can find different boundary theories. A semiclassical choice reproduces the partition function (\ref{3DRes}) with corrections resulting from non-classical backgrounds. The boundary theories can again be interpreted as describing the embedding of a quantum surface into 3D quantum flat space. In particular, \cite{qlhA} reveals a connection of these boundary theories to restricted solid-on-solid (RSOS) models, which are statistical models describing the growth of surfaces in (flat) 3D space. \\[1mm]

In this work we will consider the question of how many of these results from three-dimensional  gravity can be extended to four dimensions. The main difficulty  is that 4D gravity is, in contrast to its 3D counterpart, {\it  not} a topological theory: it features propagating curvature degrees of freedom.  We will, however, concentrate on ``the flat sector" of 4D gravity, that is, we will consider boundary metrics that induce a flat solution. In $(2+1)$D this applies to all possible boundary metrics, in $(3+1)$D this constrains two out of the six  metric degrees of freedom per (boundary) point. 

The ``flat sector" does not allow for interesting dynamics in  the bulk. However, the boundary dynamics is at least as rich as that of 3D gravity. Out of the four degrees of freedom parametrizing the boundary metrics for the flat sector, three account for diffeomorphisms along the boundary. The remaining degree of freedom describes deformations of the boundary normal to itself and captures boundary gravitons. 

We will again seek a boundary theory for these boundary gravitons. As in $(2+1)$D we will aim to extract an effective action for the lengths of a set of geodesics, stretching from the boundary to a central point or central axis. We will use Regge calculus to find this effective action. The 4D Regge action, evaluated on solutions, is not generically invariant under changes of the bulk triangulation. Invariance does hold, however, for the flat sector. This allows us to work with the coarsest bulk triangulation consistent with the continuum limit of the boundary. To make the computations feasible we will work with linearized Regge calculus and work with the closest possible analogue background solution to the one used in $(2+1)$D, which is a solid hyper-torus. 

The resulting effective action  for the geodesic lengths will be surprisingly similar to the one found in $(2+1)$D: the action is again local in the large radius limit; it has a degenerate kinetic term; the quadratic form describing this kinetic term can again be obtained from the trace-reversed extrinsic curvature tensor; and the boundary field is coupled to the boundary metric, again, via the boundary Ricci scalar. 

As mentioned above, the on-shell action is invariant under changes of the bulk triangulation for the flat sector.  However, in contrast to the 3D case one cannot find a {\it local} path integral measure such that the one-loop partition function is also invariant on the flat sector \cite{ReggeMeasure1,ReggeMeasure2}.\footnote{This holds even if one restricts to a class of triangulation changes that preserve flatness, in a particular sense.}  Nevertheless, there is one triangulation invariant feature: the degenerate kinetic term for the boundary theory leads to singularities for the one-loop correction. These singularities appeared for the 3D theory, and, in fact, completely characterized the 3D one-loop determinant. In 4D the singularitites will remain for finer triangulations because we are working with an effective radial action on the flat sector, which is triangulation invariant. Thus, again these singularities are a robust feature of the calculation.  

 In section \ref{discussion}, we briefly discuss an alternative dynamics for the kinematical variables of Regge calculus  which would allow one to completely calculate the one-loop partition function. This alternative choice imposes sharply a flat 4D space time. A quantum model with such a dynamics, and in particular a triangulation invariant path integral measure, has been constructed by Baratin and Freidel in \cite{BaratinFreidel4D}.  This invariant measure makes it possible to calculate the one-loop partition function using the coarsest possible bulk triangulation.\footnote{This partition function will have support only on boundary metrics leading to flat solutions.}  However, in this work we will focus on ``gravitational" Regge calculus, as we  plan to extend our considerations to include solutions with curvature in future work. In fact, we will find hints of an asymptotic regime in which the on-shell action has quite a simple structure, including for solutions with curvature. \\[1mm]

The paper is structured as follows: section \ref{Section:Regge}  introduces Regge calculus and the one-loop approximation to the path integral built out of this theory. Regge calculus allows us to explicitly compute the boundary dual theory for the flat sector of four-dimensional gravity. In section \ref{Locality} we establish why this boundary theory will always be local for spacetimes with the topology of a ball. In this paper, our efforts will be focused on another topology, that of the solid hyper-torus, introduced in \ref{HyperTorus} and triangulated in \ref{Triangulation}, and in this example we find that the non-localities are suppressed at large radius. Section \ref{Triangulation} also introduces the discrete Fourier transform that allows the results of the paper to be calculable. 

To carry out explicit computations it is necessary to linearize the theory around the hyper-torus background. Section \ref{ZeroFirst} carries this out for the zeroth and first order in length fluctuations, while section \ref{SecondOutline} outlines the steps that are necessary to carry this out to second order. The second order computations are begun by treating the second order radial effective action in section \ref{ComputeSecondEffective}.  This section mainly introduces some important changes of variables and definitions for what follows. Sections \ref{Effectivek2up} and \ref{sec:low} perform the main work of the computation, focusing on the higher order, and lowest lying modes, respectively. The lowest order modes are split off because they correspond to diffeomorphism symmetries that must be treated with care. The complete second order Hamilton-Jacobi functional is assembled in section \ref{section:HJF} and the resulting boundary field theory is detailed in section \ref{MainResult}. These two sections constitute the main results of the paper. Section \ref{Lowmodes} returns to the low lying diffeomorphism modes and studies how they impact the Hamilton-Jacobi functional.

Section \ref{sec:singular} takes up singularities of the Hessian not related to diffeomorphism gauge symmetries. These results are then used in section \ref{sec:oneloop} to compute the one-loop corrections to the path integral. The discussion, section \ref{discussion}, summarizes the results of the paper and gives several directions in which this work could be productively extended.

\section{Regge calculus}\label{Section:Regge}

In Regge calculus \cite{Regge} one replaces the continuous metric field on a smooth manifold with an assignment of length variables $l_e$ to the edges $e$ of a triangulation ${\cal T}$. The length variables specify a piecewise flat and linear geometry for the triangulation ${\cal T}$. For triangulations with boundary, the solutions of the theory are determined by varying the Regge  action with the appropriate boundary term. When the edge lengths on the boundary are fixed, the appropriate boundary term is the Hartle-Sorkin term \cite{HS}, which is a discretization of the Gibbons-Hawking-York term. The Regge action is
\be\label{Regge action}
-8\pi G S_R [l_e] = \sum_{t\in \cal{T}^\circ} \, A_t (l_e) \epsilon_t(l_e)+ \sum_{t\in \partial \cal{T}} \, A_t (l_e) \psi_t (l_e),
\ee
where $\cal{T}^\circ$ denotes the bulk of the triangulation $\cal{T}$, $\partial \cal{T}$ its boundary, and $A_t$ is the area of the triangle $t$. 
The bulk and boundary deficit angles, which specify the intrinsic curvature of the bulk and the extrinsic curvature of the boundary respectively,  are defined by 
\be
\epsilon_t(l_{e'}) = 2\pi - \sum_{\sigma \supset t} \theta_t^\sigma(l_{e'}) , \q \text{and} \q \psi_t(l_{e'}) = \pi - \sum_{\sigma \supset t} \theta_t^\sigma(l_{e'})  .
\ee
Here $\theta_t^\sigma$ is the interior dihedral angle in the 4-simplex $\sigma$ at the triangle $t$.

With the Hartle-Sorkin term, the Regge action is additive under gluing of two triangulations along their boundaries. 
  Varying the action with respect to the bulk edge lengths one has the equations of motion
\ba
\sum_{t\supset e } \frac{\partial A_t}{\partial l_e} \epsilon_t(l_{e'}) \,=\, 0  .
\ea
These are  discretizations of the Einstein equations for gravity. Here, as in the continuum case, where the variation of the Ricci tensor yields a total divergence, the variation of the curvature---given by the deficit angles $\epsilon_t$---also vanishes. This is due to the Schl\"afli identity  \cite{Schlafli:1858} (for a modern symplectic proof see \cite{HaggardSch:2015}),
\ba
\sum_{t\in \sigma} A_t \delta \theta_t^\sigma \,=\, 0  .
\ea

The solutions to these equations may not be unique. In particular, if the solution is flat, that is, if all deficit angles vanish, there will be a four-parameter gauge freedom for every bulk vertex \cite{Williams1,Williams2,DittrichReview08}. This gauge freedom is a remnant of the diffeomorphism symmetry of the continuum. In the Regge setup this gauge freedom can be understood as follows:  given a flat geometry with boundary, which is triangulated and thus also a piecewise linear and flat space, we can obtain a Regge solution by triangulating the bulk of this geometry. The edge lengths for this triangulation are induced by the flat geometry.    To determine the geometric data of the triangulation  we have to choose the positions for the bulk vertices inside the given flat geometry. Changing these positions  changes the lengths of the adjacent edges, that is, the bulk variables are changed without changing the flatness of the solutions and without affecting the boundary data.  Thus, also the extrinsic boundary angles are not changed, and the Regge action, which on flat solutions only contributes a boundary term, is unchanged. This defines a diffeomorphism gauge symmetry, and due to the interpretation outlined above, this symmetry is also known as a vertex translation symmetry. 

This symmetry is generically broken for solutions with curvature \cite{BahrDittrich09a,Proceeding09}.\footnote{This is true if one uses flat simplices. There is also a Regge action for homogeneously curved simplices \cite{Improve, NewRegge, Haggard:15,Haggard:16,Haggard:162}, which is particularly appropriate in the presence of a cosmological constant. The vertex translation symmetries are then present for solutions describing a homogeneously curved spacetime.}  The Hessian  evaluated on a flat solution will have, in general, four null eigenvalues for each bulk vertex. Turning on curvature, for instance by changing the boundary data, these eigenvalues will no longer vanish and will scale with the amount of curvature, as determined by the deficit angles \cite{DittrichReview08,BahrDittrich09a}.  A similar effect arises in the presence of torsion \cite{AsanteDittrichHaggard18}.

Vertex translation symmetries for flat solutions imply that Regge calculus, linearized around a flat background, will also have these symmetries. Despite the fact that time evolution in Regge calculus proceeds in discrete steps, and may even change the number of degrees of freedom, one can perform a space-time split, and perform a canonical analysis \cite{DittHoehn1,DittHoehn2,DittHoehn3}. In this analysis, the linearized theory has (linearized) first order constraints, in perfect correspondence to the continuum Hamiltonian and diffeomorphism constraints. 

Vertex translation symmetries are also linked to triangulation invariance \cite{Improve, BahrDittStein,DittrichReview12}. When evaluated on  flat solutions  (with boundaries), the action does not depend on the choice of bulk triangulation, even when the theory is linearized around a flat background. This is not the case for solutions involving curvature---even if one is considering the linearized theory \cite{ReggeMeasure1}. In summary, flat solutions in Regge calculus showcase diffeomorphism symmetry and triangulation invariance; this remains true even for homogeneously curved solutions, provided one chooses to work with homogeneously curved building blocks \cite{NewRegge}. 

Does this invariance extend to the quantum theory?  We will argue that it does not, but that there are still interesting properties of the theory that are invariant. Consider the path integral for the linearized Regge action, i.e. a one-loop approximation of the full theory. The path integral is 
\ba\label{pi01}
{\cal Z}_1(L,\ell_{\rm bdry})\,=\,  \int \mu(L)  \prod_{e\in \cal{T}^\circ} d\ell_e\, \,\,   \exp\left(-S_R^{[2]}(L,\ell) \right),
\ea
where $S_R^{[2]}$ is the expansion of the Regge action to second order in the fluctuations $\ell_e$, defined by $l_e=L_e+\ell_e$, with $L_e$ denoting the length of the edge $e$ in a flat background solution.  To evaluate the path integral we integrate over the fluctuation variables $\ell_e$ associated to the bulk edges $e \in \cal{T}^\circ$.  The integration measure is defined by $\mu(L)$, which, in this approximation, is a function of the background lengths only. 

The path integral is ill-defined for two reasons. The first reason is the diffeomorphism symmetry, which leads to non-compact gauge orbits. As in \cite{BaratinFreidel3D,BaratinFreidel4D,BonzomDittrich}, we will identify a measure over these gauge orbits and split off these infinite integrals. (This is equivalent to gauge fixing and inserting a Faddeev-Popov determinant.) The second reason is known as the conformal mode problem: the gravitational action is unbounded from below due to the kinetic term from the conformal mode. We will treat this problem by formally rotating the sign of this mode \cite{ConformalMode}. 

Given that the path integral for the linearized theory features a notion of diffeomorphism symmetry, one can ask if there is a choice for the measure factor $\mu(L)$ that would make the partition function invariant under triangulation changes \cite{ReggeMeasure1}.  One can even consider a subset of local triangulation changes that leave the classical theory invariant.\footnote{
These would be the $1-5$ and $2-4$ Pachner moves and their inverses. A $x$-$( 6-x)$ Pachner move replaces a complex of $x$ four-simplices with a complex of $(6-x)$ four-simplices \cite{Pachner}. Both complexes have the same boundary triangulation. The boundary triangulation of the $1-5$ and $2-4$ Pachner moves only allow for a flat bulk solution. This is the reason why the full and linearized Regge actions (evaluated on the corresponding solutions) are invariant under these Pachner moves. By contrast the $3-3$ move complex also allows for curvature. Neither the full nor the linearized Regge actions are invariant under this move, if one considers a solution with curvature \cite{ReggeMeasure1}. Every (bulk) triangulation change can be obtained by a sequence of the full set of Pachner moves.
}
There is, however, even with this restriction, no {\it local} choice of measure that would make the one-loop partition function invariant \cite{ReggeMeasure2}.

Not having such an invariant measure (either local or non-local) at hand, the one-loop correction will depend on the choice of triangulation, even if we consider boundary data inducing a flat solution. However, here we will be interested in singularities for the one-loop correction, which result from zero's in the determinant of the Hessian of the Regge action (after removing the zero's resulting from the gauge symmetries). The existence of these zero's {\em is} independent of the choice of bulk triangulation. As we will see, they result from a degeneracy of the kinetic term for the dual boundary field theory. 

 As we do not have a triangulation invariant measure at our disposal, we will not further specify the `bare' measure $\mu(L)$---it will appear as a multiplicative factor for the one-loop correction.  See  \cite{hamber, menotti, ReggeMeasure1,ReggeMeasure2} for suggestions for this measure term, including non-local constructions.

As discussed at the end of the introduction, we can also employ an alternative theory, constructed in \cite{BaratinFreidel4D}, which describes (quantum) flat space. This theory is  topological, that is, (bulk) triangulation invariant, and includes, in particular, an invariant measure term.

\section{On the locality of the effective boundary field theory}\label{Locality}


\newcommand\pgfmathsinandcos[3]{%
  \pgfmathsetmacro#1{sin(#3)}%
  \pgfmathsetmacro#2{cos(#3)}%
}
\newcommand\LongitudePlane[3][current plane]{%
  \pgfmathsinandcos\sinEl\cosEl{#2} 
  \pgfmathsinandcos\sint\cost{#3} 
  \tikzset{#1/.style={cm={\cost,\sint*\sinEl,0,\cosEl,(0,0)}}}
}

\newcommand\LatitudePlane[3][current plane]{%
  \pgfmathsinandcos\sinEl\cosEl{#2} 
  \pgfmathsinandcos\sint\cost{#3} 
  \pgfmathsetmacro\yshift{\cosEl*\sint}
  \tikzset{#1/.style={cm={\cost,0,0,\cost*\sinEl,(0,\yshift)}}} %
}
\newcommand\NewLatitudePlane[4][current plane]{%
  \pgfmathsinandcos\sinEl\cosEl{#3} 
  \pgfmathsinandcos\sint\cost{#4} 
  \pgfmathsetmacro\yshift{#2*\cosEl*\sint}
  \tikzset{#1/.style={cm={\cost,0,0,\cost*\sinEl,(0,\yshift)}}} %
}
\newcommand\DrawLongitudeCircle[2][1]{
  \LongitudePlane{\angEl}{#2}
  \tikzset{current plane/.prefix style={scale=#1}}
  \pgfmathsetmacro\angVis{atan(sin(#2)*cos(\angEl)/sin(\angEl))} %
  \draw[current plane] (\angVis:1) arc (\angVis:\angVis+180:1);
}
\newcommand\DrawLatitudeCircle[2][1]{
  \LatitudePlane{\angEl}{#2}
  \tikzset{current plane/.prefix style={scale=#1}}
  \pgfmathsetmacro\sinVis{sin(#2)/cos(#2)*sin(\angEl)/cos(\angEl)}
  \pgfmathsetmacro\angVis{asin(min(1,max(\sinVis,-1)))}
  \draw[current plane] (\angVis:1) arc (\angVis:-\angVis-180:1);
}
\newcommand\DrawLongitudeCirclered[2][1]{
    \LongitudePlane{\angEl}{#2}
    \tikzset{current plane/.estyle={cm={\cost,\sint*\sinEl,0,\cosEl,(0,0)},scale=#1}}
    \pgfmathsetmacro\angVis{atan(sin(#2)*cos(\angEl)/sin(\angEl))} %
    \draw[current plane,red!80,thick] (90:1) arc (90:180:1);   
}
\newcommand\DrawLatitudeCirclered[2][1]{
    \LatitudePlane{\angEl}{#2}
    \tikzset{current plane/.prefix style={scale=#1}}
    \pgfmathsetmacro\sinVis{sin(#2)/cos(#2)*sin(\angEl)/cos(\angEl)}
    \pgfmathsetmacro\angVis{asin(min(1,max(\sinVis,-1)))}
    \draw[current plane,red!80,thick] (\angPhiTwo:1) node[below right] {$$} arc (\angPhiTwo:\angPhiOne:1) node[below left] {$$}; 
}

\newcommand\DrawLatitudeCircleMod[2][1]{
    \LatitudePlane{\angEl}{#2}
    \tikzset{current plane/.prefix style={scale=#1}}
    \pgfmathsetmacro\sinVis{sin(#2)/cos(#2)*sin(\angEl)/cos(\angEl)}
    \pgfmathsetmacro\angVis{asin(min(1,max(\sinVis,-1)))}
    \draw[current plane]  (\angVis:1) arc (\angVis:\angPhiTwo:1) (\angPhiOne:1) arc (\angPhiOne:-\angVis-180:1) ; 
}

\newcommand\DrawLatitudeArc[4][black]{
  \LatitudePlane{\angEl}{#2}
  \tikzset{current plane/.prefix style={scale=1}}
  \pgfmathsetmacro\sinVis{sin(#2)/cos(#2)*sin(\angEl)/cos(\angEl)}
  \pgfmathsetmacro\angVis{asin(min(1,max(\sinVis,-1)))}
  \pgfmathsetmacro\angA{max(min(\angVis,#3),-\angVis-180)} %
  \pgfmathsetmacro\angB{min(\angVis,#4)} %
  \draw[current plane,#1,opacity=0.4] (#3:\R) arc (#3:#4:\R);
  \draw[current plane,#1] (\angA:\R) arc (\angA:\angB:\R);
}

\tikzset{%
  >=latex, 
  inner sep=0pt,%
  outer sep=2pt,%
  mark coordinate/.style={inner sep=0pt,outer sep=0pt,minimum size=3pt,
    fill=black,circle}%
}

\begin{figure}

\begin{tikzpicture}[scale = 0.5] 
\def\R{4 } 
\def\angEl{20} 
\def\angAz{-20} 
\def\angPhiOne{-120} 
\def\angPhiTwo{-40} 
\def\angBeta{10} 
\filldraw[fill=white] (0,0) circle (\R);

\foreach \t in {-55,-25,0} { \DrawLatitudeCircle[\R]{\t} }
\foreach \t in {-40,30,60} { \DrawLongitudeCircle[\R]{\t} }

\DrawLatitudeCircleMod[\R]{30}
\DrawLatitudeCircleMod[\R]{60}

\pgfmathsetmacro\H{\R*cos(\angEl)} 
\pgfmathsetmacro\K{\R*cos(\angAz)} 
\coordinate (O) at (0,0);
\node[circle,draw,black,scale=0.3] at (0,0) {};
\draw[above left] node at (0,0){O};
\coordinate[mark coordinate] (N) at (0,\H);
\coordinate[mark coordinate] (S) at (0,-\H);

\LongitudePlane[angle]{\angEl}{-120};

\NewLatitudePlane[radiiA]{\R}{\angEl}{60};
\path[radiiA] (15:\R) coordinate (A1);
\path[radiiA] (-40:\R) coordinate (A2);
\path[radiiA] (-120:\R) coordinate (A3);
\path[radiiA] (-150:\R) coordinate (A4);

\NewLatitudePlane[radiiB]{\R}{\angEl}{30};
\path[radiiB] (0:\R) coordinate (B1);
\path[radiiB] (-40:\R) coordinate (B2);
\path[radiiB] (-120:\R) coordinate (B3);
\path[radiiB] (-150:\R) coordinate (B4);
\path[radiiB] (-180:\R) coordinate (B5);

\NewLatitudePlane[radiiC]{\R}{\angEl}{00};
\path[radiiC] (0:\R) coordinate (C1);
\path[radiiC] (-40:\R) coordinate (C2);
\path[radiiC] (-65:\R) coordinate (C3);
\path[radiiC] (-95:\R) coordinate (C4);
\path[radiiC] (-120:\R) coordinate (C5);
\path[radiiC] (-150:\R) coordinate (C6);
\path[radiiC] (-185:\R) coordinate (C7);

\NewLatitudePlane[radiiD]{\R}{\angEl}{-25};
\path[radiiD] (0:\R) coordinate (D1);
\path[radiiD] (-40:\R) coordinate (D2);
\path[radiiD] (-65:\R) coordinate (D3);
\path[radiiD] (-95:\R) coordinate (D4);
\path[radiiD] (-120:\R) coordinate (D5);
\path[radiiD] (-150:\R) coordinate (D6);
\path[radiiD] (-180:\R) coordinate (D7);

\NewLatitudePlane[radiiE]{\R}{\angEl}{-55};
\path[radiiE] (-65:\R) coordinate (E1);
\path[radiiE] (-95:\R) coordinate (E2);
\path[radiiE] (-120:\R) coordinate (E3);
\path[radiiE] (-150:\R) coordinate (E4);


\filldraw[gray!20] (C2) to [bend left = 10] (C5) to [bend left = 22](N) to [bend left = 35] (C2);

\foreach \t in {N,C2,C3,C4,C5,B2,B3,A2,A3} {\draw[-, thick , dashed] (O)--(\t); }
\foreach \t in {N,C2,C5} {\draw[-, dotted ,thick] (O)--(\t); }

\LongitudePlane[angle]{\angEl}{-95};
\draw[angle,-] (0:\R) arc (0:-65:4);
\LongitudePlane[angle]{\angEl}{-65};
\draw[angle,-] (0:\R) arc (0:-65:4);

\DrawLongitudeCirclered[\R]{180+\angPhiOne}
\DrawLongitudeCirclered[\R]{180+\angPhiTwo}
\DrawLatitudeCirclered[\R]{0}

\draw[gray!70,thick] (B2) to [bend right = 25] (A1);
\draw[gray!70,thick] (B4) to [bend right = 0] (A3);
\draw[gray!70,thick] (B5) to [bend left = 10] (A4);

\draw[gray!70,thick] (C2) to [bend right= 10] (B1);
\draw[gray!70,thick] (C6) to [bend left= 10] (B3);
\draw[gray!70,thick] (C7) to [bend left= 15] (B4);

\draw[gray!70,thick] (D2) to [bend right= 15] (C1);
\draw[gray!70,thick] (D3) to [bend right= 15] (C2);
\draw[gray!70,thick] (D4) to [bend right= 12] (C3);
\draw[gray!70,thick] (D5) to [bend right= 10] (C4);
\draw[gray!70,thick] (D6) to [bend right= 5] (C5);
\draw[gray!70,thick] (D7) to [bend left= 5] (C6);

\draw[gray!70,thick] (E1) to [bend right = 10] (D2);
\draw[gray!70,thick] (E2) to [bend right = 10] (D3);
\draw[gray!70,thick] (E3) to [bend right = 0] (D4);
\draw[gray!70,thick] (E4) to [bend left = 5] (D5);

\end{tikzpicture} 

\caption{A triangulation of a ball shaped region. All bulk edges are radial, going from the boundary to a central vertex. The Regge action for this triangulation can be understood as a boundary action for the radial edge lengths.}
 \label{BallTr}
\end{figure}
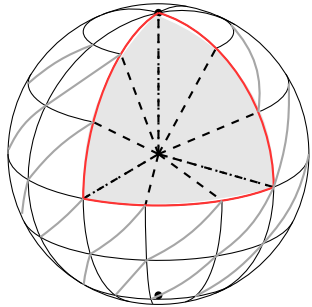

  Having covered the basics of Regge calculus we will now explain how to define the dynamics for a boundary field given by the lengths of geodesics going from the boundary to some central point. Using Regge calculus we will see that the action describing this dynamics is local for a ball-shaped region---at least if we restrict to boundary metrics that induce flat solutions. Here we define an action to be local, if it couples only variables associated to building blocks that are neighbors of some finite degree, e.g. next-to-nearest neighbors. This translates, in the continuum limit, to having only finite order differential operators appearing in the action. What follows applies to any bulk dimension $D \geq 3$. For $D=3$ all boundary metrics induce a flat solution, whereas for $D=4$ this holds only for a subset.
  
The Hamilton-Jacobi functional, that is, the on-shell action, is invariant under bulk triangulation changes for boundary metrics that induce a flat solution. Thus, restricting to these boundary metrics, we can work with any bulk triangulation. For a ball-shaped region we can in particular choose a triangulation that has only one bulk vertex, but still allow for arbitrarily many vertices on the boundary. Thus all the bulk edges are radial and go from a boundary vertex to the bulk vertex, see Fig. \ref{BallTr}. The geometry defined by Regge calculus is piecewise linear and flat.\footnote{This holds in the original version \cite{Regge} of Regge calculus. Note that one can also work with homogeneously curved building blocks \cite{Improve, NewRegge, Haggard:15,Haggard:16,Haggard:162}, which allows generalization of this argument to (Regge) gravity with a cosmological constant.} Given a flat solution the edges will therefore coincide with geodesics. Even off-shell we can define geodesics inside a given building block that are straight lines in the flat geometry of the given building block, and go from the boundary to the central vertex. We can place these geodesics arbitrarily close to the edges of the block and, thus, the geodesics' lengths will approximate arbitrarily well the lengths of the edges.

The Regge action for such a triangulation will be a function of the boundary edge lengths and the radial bulk edge lengths. The latter can be identified with a boundary field, giving the geodesic radial distance of the boundary to the central vertex. Thus the Regge action itself defines an ``effective" theory for the geodesic radial distance.

As explained in section \ref{Section:Regge}, the displacement of the central vertex is a remnant of diffeomorphism symmetry. It acts as a gauge symmetry on the boundary field, at least from the perspective of the gravitational partition function.  It can also be seen as a global symmetry from the perspective of the boundary theory---in fact, it provides only global symmetry parameters. However, to regain the partition function of gravity, we have to treat this symmetry as a gauge symmetry. 

The effective theory defined by the Regge action is local in the following sense: for two length variables associated to two edges $e$ and $e'$ to couple to each other, $e$ and $e'$ must both be included in at least one $D$-simplex. Translated to the boundary, this means that two radial length variables associated to two boundary vertices $v$ and $v'$ can only be coupled to each other if $v$ and $v'$ are both included in at least one boundary $(D-1)$-simplex. Thus, the Regge action is also local when interpreted as a boundary theory. This locality continues to hold for the coupling of the boundary field to the boundary lengths. 
 
 For ball-shaped regions we therefore obtain a local boundary theory. This does not necessarily hold for other topologies, indeed,  for a solid hyper-torus we also obtain non-local terms. However, direct computation shows that these are suppressed in the limit of large radii.  \\

\section{The background space time}\label{HyperTorus}

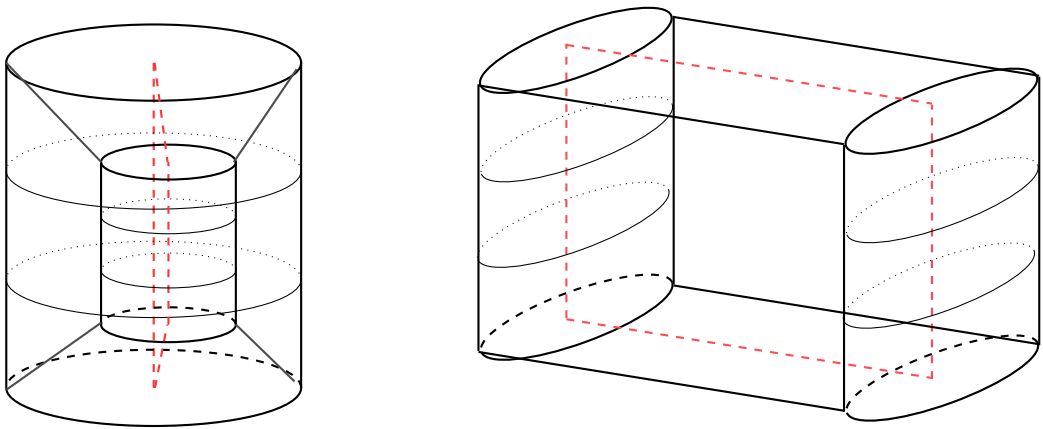
\begin{figure}[ht!]
\def\a{45}
\tdplotsetmaincoords{75}{\a}
\begin{tikzpicture}[scale=5.6,tdplot_main_coords,axis/.style={->,red!80,very thin},curve/.style={black,thin}]
\begin{scope}
\def\radius{.35}
\def\axissize{.5}
\def\th{.8}
\draw[axis,-,dashed,thick] (0,0,0)node[below] (A) {} -- (0,0,\th)node[above] (B) {} ;

\tdplotsinandcos{\sintheta}{\costheta}{0}
\tdplotdrawarc[curve,thick]{(0,0,\th)}{\radius*\costheta}{\a-360}{\a}{}{}
\foreach \height in {{\th/3},{2*\th/3}}{
\tdplotdrawarc[curve,solid]{(0,0,\height)}{\radius*\costheta}{\a-180}{\a}{}{}
\tdplotdrawarc[curve,dotted]{(0,0,\height)}{\radius*\costheta}{\a-360}{\a-180}{}{}
    }
\tdplotdrawarc[curve,thick,dashed]{(0,0,0)}{\radius*\costheta}{\a}{\a+180}{}{}
\tdplotdrawarc[curve,thick]{(0,0,0)}{\radius*\costheta}{\a}{\a-180}{}{}
\tdplotsinandcos{\sintheta}{\costheta}{\a}
\draw[thick] (\radius*\costheta,\radius*\sintheta,0) node[ ] (C) {} -- (\radius*\costheta,\radius*\sintheta,\th) node[ ] (D) {};
\tdplotsinandcos{\sintheta}{\costheta}{\a+180}
\draw[thick] (\radius*\costheta,\radius*\sintheta,0) node[below left] (E) {} -- (\radius*\costheta,\radius*\sintheta,\th) node[above left] (F) {};

\end{scope}
\begin{scope}[xshift=.035cm ,yshift = .15cm]
\def\radius{.16}
\def\axissize{.3}
\def\th{.4}
\draw[axis,-,dashed,thick] (0,0,0) node[above] (a) {} -- (0,0,\th) node[below] (b) {};
\node[red,right] at (0.00,0,0) {} ;
\node[red,right] at (0.00,0,\th) {} ;
\tdplotsinandcos{\sintheta}{\costheta}{0}
\tdplotdrawarc[curve,thick]{(0,0,\th)}{\radius*\costheta}{\a-360}{\a}{}{}
\foreach \height in {{\th/3},{2*\th/3}}{
\tdplotdrawarc[curve,solid]{(0,0,\height)}{\radius*\costheta}{\a-180}{\a}{}{}
\tdplotdrawarc[curve,dotted]{(0,0,\height)}{\radius*\costheta}{\a-360}{\a-180}{}{}
    }
\tdplotdrawarc[curve,thick,dashed]{(0,0,0)}{\radius*\costheta}{\a}{\a+180}{}{}
\tdplotdrawarc[curve,thick]{(0,0,0)}{\radius*\costheta}{\a}{\a-180}{}{}
\tdplotsinandcos{\sintheta}{\costheta}{\a}
\draw[thick] (\radius*\costheta,\radius*\sintheta,0) node[above left] (c) {} -- (\radius*\costheta,\radius*\sintheta,\th) node[below left ] (d) {};
\tdplotsinandcos{\sintheta}{\costheta}{\a+180}
\draw[thick] (\radius*\costheta,\radius*\sintheta,0) node[above right] (e) {} -- (\radius*\costheta,\radius*\sintheta,\th) node[below right] (f) {} ;
\end{scope}
\draw[thick,red!80,dashed] (A)--(a) (B)--(b) ;
\draw[black!70,thick] (C)--(c) (D)--(d) (E)--(e) (F)--(f);
\end{tikzpicture} \hspace{2cm}
\begin{tikzpicture}[scale=5.4,tdplot_main_coords,axis/.style={->,red!70,very thin},curve/.style={black,thin}]
\begin{scope}
\def\radius{.25}
\def\axissize{.5}
\def\th{.7}
\draw[axis,-,dashed,thick] node[ left] (A) {}(0,0,0) -- (0,0,\th)node[left] (B) {};

\tdplotsinandcos{\sintheta}{\costheta}{0}
\tdplotdrawarc[curve,thick,rotate=20]{(.35,0,\th)}{\radius*\costheta}{\a-360}{\a}{}{}
\foreach \height in {{\th/3+0.01},{2*\th/3}}{
\tdplotdrawarc[curve,solid,rotate=20]{(0.53*\height,0,\height)}{\radius*\costheta}{\a-180}{\a}{}{}
\tdplotdrawarc[curve,dotted,rotate=20]{(0.53*\height,0,\height)}{\radius*\costheta}{\a-360}{\a-180}{}{}
    }
\tdplotdrawarc[curve,thick,dashed,rotate=20]{(0.035,0,0)}{\radius*\costheta}{\a}{\a+180}{}{}
\tdplotdrawarc[curve,thick,rotate=20]{(0.035,0,0)}{\radius*\costheta}{\a}{\a-180}{}{}
\tdplotsinandcos{\sintheta}{\costheta}{\a}
\draw[thick]  (\radius*\costheta+0.02,\radius*\sintheta,0.09) node[left ] (C) {}  -- (\radius*\costheta+0.02,\radius*\sintheta,\th+0.075) node[left] (D) {};
\tdplotsinandcos{\sintheta}{\costheta}{\a-150}
\draw[thick]  (\radius*\costheta,\radius*\sintheta,-0.05) node[left] (E) {} -- (\radius*\costheta,\radius*\sintheta,\th-0.07) node[left] (F) {} ;

\end{scope}
\begin{scope}[xshift=.9cm ,yshift = -.15cm]
\def\radius{.25}
\def\axissize{.3}
\def\th{.7}
\draw[axis,-,dashed,thick] node[right] (a) {} (0,0,0) -- (0,0,\th) node[right] (b) {};

\tdplotsinandcos{\sintheta}{\costheta}{0}
\tdplotdrawarc[curve,thick,rotate=20]{(.35,0,\th)}{\radius*\costheta}{\a-360}{\a}{}{}
\foreach \height in {{\th/3+0.01},{2*\th/3}}{
\tdplotdrawarc[curve,solid,rotate=20]{(0.53*\height,0,\height)}{\radius*\costheta}{\a-180}{\a}{}{}
\tdplotdrawarc[curve,dotted,rotate=20]{(0.53*\height,0,\height)}{\radius*\costheta}{\a-360}{\a-180}{}{}
    }
\tdplotdrawarc[curve,thick,dashed,rotate=20]{(0.035,0,0)}{\radius*\costheta}{\a}{\a+180}{}{}
\tdplotdrawarc[curve,thick,rotate=20]{(0.035,0,0)}{\radius*\costheta}{\a}{\a-180}{}{}
\tdplotsinandcos{\sintheta}{\costheta}{\a}
\draw[thick] (\radius*\costheta+0.02,\radius*\sintheta,0.09) node[right] (c) {}   -- (\radius*\costheta+0.02,\radius*\sintheta,\th+0.075) node[right] (d) {};
\tdplotsinandcos{\sintheta}{\costheta}{\a-150}
\draw[thick] (\radius*\costheta,\radius*\sintheta,-0.05) node[ right] (e) {}  -- (\radius*\costheta,\radius*\sintheta,\th-0.07) node[ right] (f) {};
\end{scope}
\draw[thick,red!70,dashed] (A) -- (a) (B)--(b) ;
\draw[thick] (C) -- (c) (D)--(d) (E) -- (e) (F)--(f);
\end{tikzpicture}
\caption{Different 2D projections of a 4D hyper-clinder.  The red  dashed lines indicate the 2D axis where $r=0$.  }
\label{HyperCylinder}
\end{figure}

The background spacetime we will consider is a Euclidean signature, flat spacetime with boundary. It generalizes the 3D spacetime known as thermal spinning flat space \cite{Barnich}.  This 3D spacetime is obtained by taking a solid cylinder with radius $R$ and twisting this cylinder by an angle $\gamma$ around its axis before gluing it to a solid torus.  One often uses a ``time" coordinate $t$ along the axis of the cylinder, an angular coordinate $\theta$ that goes around the central axis, and a radial coordinate $r$. 

For our four-dimensional spacetime we  replace the ``time" coordinate $t$  with two coordinates $y$ and $z$. The solid hyper-cylinder is  $D\times [0,\alpha]\times [0,\beta]$, where $D$ is the two-dimensional disk and it has a two-dimensional central  axis $[0,\alpha] \times [0,\beta]$ coordinatized by $y$ and $z$, see Fig. \ref{HyperCylinder}.  To get a spacetime with one boundary component we glue the cylinder twice. We first identify the boundaries $D\times\{0\}\times[0,\beta]$ and $D\times\{\alpha\}\times[0,\beta]$ with each other after rotating the latter by an angle $\gamma_y$. Next we identify $D\times S^1 \times \{0\}$ and $D \times S_1 \times \{\beta\}$, once again inserting a rotation by $\gamma_z$ of the disk in the second component. This gives  the spacetime 
 \ba\label{contmetric}
ds^2 =dr^2 + r^2 d\theta^2 + dy^2 + dz^2 
\ea
with $
r\in [0,R]$
and the remaining coordinates subject to the following periodic identifications 
\ba
\label{identifications}
(r,\theta, y, z) &\sim& (r,\theta+2\pi,y,z), \nn\\
(r,\theta, y, z) &\sim& (r,\theta+\gamma_y,y + \alpha ,z ),   \nn\\
\text{and} \quad (r,\theta,y,z) &\sim& (r,\theta+\gamma_z,y, z + \beta ) .
\ea

Let us evaluate the Einstein action with the Gibbons-Hawking-York boundary term
\ba
S\,=\,  -\frac{1}{16\pi G} \int \sqrt{g} R \, d^4x - \frac{1}{8\pi G} \int \sqrt{h} K \, d^3x
\ea
on this spacetime. There is only a contribution from the boundary term. The extrinsic curvature tensor of the $r=R$ hypersurface is $K_{ab}=\text{diag}(R,0,0)$ and the trace is given by $K=\tfrac{1}{R}$. Together with $\sqrt{h}=R$ this leads to a boundary term which is proportional to the area of the hypersurface at $R=1$:
\ba
S\,=\,-\frac{ \alpha \beta}{4G}  .
\ea
Note that the twist angles $\gamma_y$ and $\gamma_z$ do not appear in the classical background action. The one-loop correction will depend on these angles.

\section{Hyper-torus Triangulation and Discrete Fourier Transform}\label{Triangulation}

As discussed in section \ref{Section:Regge}, the Regge action evaluated on flat solutions will be bulk triangulation independent. This allows us to choose a very coarse bulk triangulation. However, we also want to take the continuum limit on the boundary and will choose a sufficiently general and regular triangulation to achieve this limit. 

The spacetime under consideration has the topology of a solid three-torus, i.e. $D\times S^1\times S^1$ where $D$ is a disk and $S^1$ the circle.  We cut this three-torus perpendicular  to the two $S^1$-directions, that is, along three-planes with  fixed $y$- and $z$-coordinates.  (Care must be taken with the twist parameters if these pieces are to be re-glued.) Repeatedly cutting in this manner we produce $N_y \times N_z$ building blocks with topology $D\times [0,1]\times [0,1]$. 

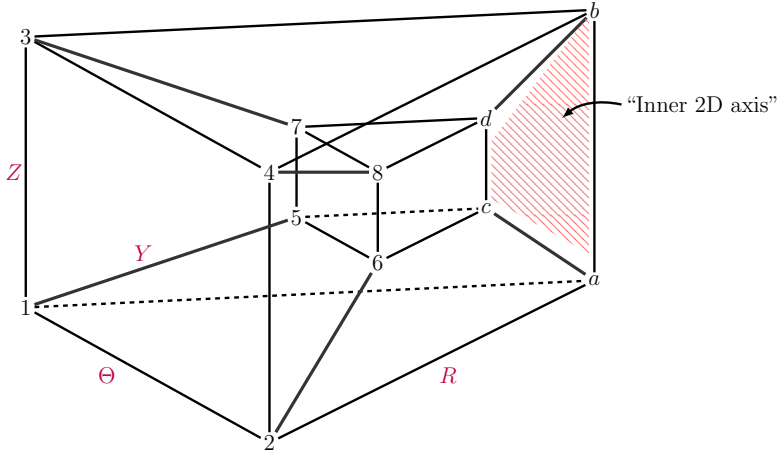
\begin{figure}[ht!]
 \scalebox{0.6}
 {
\begin{tikzpicture}[xscale=1.2]
\tikzstyle{every node}=[font=\Large]
\tikzstyle{edge1} = [draw,line width=1.5pt,-,black]
\tikzstyle{edge2} = [draw,line width=2pt,-,black!80]
\tikzstyle{edge3} = [draw,line width=2pt,-,blue!50]
\begin{scope}[xshift=2cm,yshift=4cm]
\node (v6) at (0,0) {$6$};
\node (v5) at (-1.5,1) {$5$};
\node (v7) at (-1.5,3) {$7$};
\node (v8) at (0,2) {$8$};
\node (vd) at (2,3.2) {$d$};
\node (vc) at (2,1.2) {$c$};

\draw[edge1]  (v6) -- (v5) -- (v7) -- (v8) -- (vd) -- (vc) -- (v6) -- (v8);
\draw[edge1]  (v7)--(vd);
\draw[edge1,dashed] (v5) -- (vc);
\end{scope}
\begin{scope}[scale=3.]
\node (v2) at (0,0) {$2$};
\node (v1) at (-1.5,1) {$1$};
\node (v3) at (-1.5,3) {$3$};
\node (v4) at (0,2) {$4$};
\node (vb) at (2,3.2) {$b$};
\node (va) at (2,1.2) {$a$};
\draw[edge1] (v2) -- (v1) -- (v3) -- (v4) -- (vb) -- (va) -- (v2) -- (v4);
\draw[edge1] (v3)--(vb);
\draw[edge1,dashed] (v1) -- (va);
{\color{purple}
\node at (-1,0.5) {$\Theta$};
\node[left] at (-1.5,2) {$Z$};
\node[left] at (-0.7,1.4) {$Y$};
\node at (1.1,0.5) {$R$};
}
\end{scope}
\draw[edge2] (v2) -- (v6);
\draw[edge2] (v1) -- (v5);
\draw[edge2] (v3) -- (v7);
\draw[edge2] (v4) -- (v8);
\draw[edge2] (vd) -- (vb);
\draw[edge2] (vc) -- (va);
\fill[pattern = north west lines,pattern color=red!50] (4.1,6.7) -- (4.1,5.4) -- (5.9,4.2) -- (5.9,9.5) -- (4.1,6.7);
\draw[edge1,<-] (5.4,7.2) to[bend left] (6.5,7.5);
\node[right] at (6.5,7.5) {``Inner 2D axis"}; 
\end{tikzpicture}
}
\caption{A hyper-prism as a unit block for the discretization of the solid three-torus. The numbers $1$ to $8$ indicate vertices that are positioned on the outer boundary of the solid three-torus. The small latin letters $a$ to $d$ label vertices on the two-dimensional central axis. Table \ref{lengthTab} lists all edges of the hyper-prism, together with their background lengths. Note that we have not included diagonal and hyper-diagonal edges in this figure. } \label{hypr}
\end{figure}

These building blocks are then cut along three-planes perpendicular to the disk and along three-planes with constant angular coordinate $\theta$. All these cuts go through a ``two-dimensional axis" where the radial coordinate vanishes, $r=0$. This results in $N_\theta$ hyper-prisms, see figure \ref{hypr}, each with side lengths $\Theta,Y,Z,$ and $R$.

Each hyper-prism can be triangulated into twelve four-simplices. This introduces various diagonals and hyper-diagonals. Tables \ref{lengthTab} lists all of the edges in the triangulation using a notation where $e(xy)$ denotes the edge that connects vertex $x$ to vertex $y$. This table also collects edges into groups with common background length parameter, denoted with a capital variable $L_e$, and the associated length fluctuation variable, denoted with a lower case $\ell_e$. The total length of edge $e$ is $l_e=L_e+\ell_e$. 

In the background geometry we choose lengths for the diagonals and hyper-diagonals, see lower portions of Tables \ref{lengthTab}, such that the prism is almost everywhere flat, that is, the deficit angles are vanishing for almost all bulk triangles. Exceptions are the triangles of the inner 2D axis at $r=0$.  To have a vanishing deficit angle for these triangles we need to impose a relation between the number of hyper-prisms $N_\theta$ in one  constant $(y,z)$-slice and the background lengths $R$ and $\Theta$:
\ba
x&:=& \frac{\Theta^2}{2R^2} \,\stackrel{!}{=}\, 1-\cos \left(\frac{2\pi}{N_\theta}\right) . 
\ea
Furthermore, we have $Y \times N_y\,=\, \alpha$ and $Z\times N_z\,=\, \beta$, where $\alpha$ and $\beta$ are characteristic lengths of the continuum geometry, defined above (\ref{contmetric}).

\begin{table} \scalebox{0.95}{ 
\begin{tabular}{ |c c |c|p {1.8cm }|}
\hline
\multicolumn{4}{ |c| }{Bulk edge length  variables} \\
\hline \cline{1-4}
 \multicolumn{2} {|p {2.2cm } | }{  Edges on hyper-prism} & Length of edge(s)  & Length fluctuations \\ \hline \cline{1-4}
 $e(1a)$ & $e(5c)$ & \multirow{4}{*}{$R$}  & \multirow{4}{*}{\q $\ell_r$}  \\
  $e(2a)$ & $e(6c)$ & &\\
  $e(3b)$ & $e(7d)$ & &\\
  $e(4b)$ & $e(8d)$ & & \\ \hline
 $e(1c)$ & $e(3d)$ & \multirow{2}{*}{$\sqrt{R^2+Y^2}$}  & \multirow{2}{*}{\q $\ell_{ry}$}  \\
  $e(2c)$ & $e(4d)$ & &\\ \hline
  $e(1b)$ & $e(5d)$ & \multirow{2}{*}{$\sqrt{R^2+Z^2}$} & \multirow{2}{*}{\q $\ell_{rz}$}  \\ 
  $e(2b)$ & $e(6d)$ & &\\ \hline
 $e(1d)$ & $e(2d)$  & $\sqrt{R^2+Y^2+Z^2}$  &  \q $\ell_{r y z }$ \phantom{\Big|}  \\ \hline
 $e(ac)$ & $e(bd)$ & $Y$  & \q $\ell_{\varphi}$ \phantom{\Big|}  \\ \hline
 $e(ab)$ & $e(cd)$ & $Z$  & \q $\ell_{\zeta}$ \phantom{\Big|}   \\ \hline
$e(ad)$ & & $\sqrt{Y^2+Z^2}$ & \q $\ell_{\varphi \zeta}\phantom{\Big|}$  \\ \hline
\end{tabular}
\q
\begin{tabular}{ |c c |c|p {1.8cm }|}
\hline 
\multicolumn{4}{ |c| }{Boundary edge length  variables} \\
\hline \cline{1-4}
 \multicolumn{2} {|p {2.2cm } | }{  Edges on hyper-prism} & Length of edge(s)  & Length fluctuations \\ \hline \cline{1-4}
 $e(12)$ & $e(56)$ & \multirow{2}{*}{$\Theta$}  & \multirow{2}{*}{\q $\ell_{\theta}$} \\
 $e(34)$ & $e(78)$ & &\\ \hline
 $e(15)$ & $e(37)$ & \multirow{2}{*}{$Y$}  & \multirow{2}{*}{\q $\ell_{y}$} \\
 $e(26)$ & $e(48)$ & & \\ \hline
 $e(13)$ & $e(57)$ & \multirow{2}{*}{$Z$}  &  \multirow{2}{*}{\q $\ell_{z}$} \\ 
 $e(24)$ & $e(68)$ & & \\ \hline
 $e(16)$ & $e(38)$ & $\sqrt{\Theta^2+Y^2}$ & \q $\ell_{\theta y}$ \\ \hline
 $e(14)$ & $e(58)$ & $\sqrt{\Theta^2+Z^2}$ & \q $\ell_{\theta z}$ \\ \hline
 $e(28)$ & $e(17)$ & $\sqrt{Y^2+Z^2}$ & \q $\ell_{y z}$   \\ \hline
 $e(18)$ & & $\sqrt{\Theta^2+Y^2+Z^2}$ & \q $\ell_{\theta y z}$ \\ \hline 
\end{tabular}  }

\caption{The tables relate the length variables to the edges in the hyper-prism, which are shown in Fig. \ref{hypr}.  The left table includes all edges that are in the bulk of the solid three-torus. The right table includes all the edges that are in the boundary of the solid three-torus. }
\label{lengthTab}
\end{table}

The boundary of the solid three-torus is discretized into a regular cubical lattice, with edge-lengths $\Theta,Y,$ and $Z$. The vertices of this lattice are labelled by the set
\ba
(s_\theta,s_y,s_z) \in [0,1,\ldots, N_\theta-1]\times [ 0,1,\ldots, N_y-1]\times [0,1, \ldots, N_z-1].
\ea
The cuboids are further subdivided into  six tetrahedra, which introduces face-diagonals, body-diagonals, and a hyper-diagonal, see Table \ref{lengthTab}. All the diagonals are chosen so that there is an orientation for all edges with the following property: each coordinate of the source vertex of any given edge is smaller or equal to the corresponding coordinate of the target vertex.  (Here we use the periodic identification to imagine an infinite lattice.)
In Table \ref{lengthTab} we have listed all edges in the form $e(v_1v_2)$, where $v_i$ takes values $1$ to $8$, if it is a vertex on the $r=R$ boundary of the solid three-torus, and values $a$ to $d$ if the vertex is on the two-dimensional central axis. With the above choice of orientation $v_1$  is the source vertex and $v_2$ the target vertex of an edge $e(v_1v_2)$ appearing in the table.
We can thus associate the length fluctuation variable of a given edge to its source vertex, that is our variables on the boundary are: 
\ba\label{boundary edges}
\ell_e(s_\theta,s_y,s_z) \q \text{with}\q e \in \{\theta,y,z,\theta y, \theta z, \theta y z \}  .
\ea

Moving on to the bulk edges we consider the set of edges that have one vertex, which we choose as source, at the boundary of the solid three-torus, i.e. at $r=R$, and the other vertex on the two-dimensional axis, at $r=0$. We also associate the coordinates of the source vertex to the variables associated to these edges:
\ba\label{bulk eges}
\ell_e(s_\theta,s_y,s_z) \q \text{with}\q e\in \{r, ry,rz, ryz \}  .
\ea
We have furthermore a set of bulk variables that have only vertices at $r=0$, that is on the two-dimensional axis. This axis is topologically a two-torus, and is discretized into rectangles, which are furthermore subdivided by parallel diagonals into triangles. The vertices are parametrized by $(s_y,s_z)$. Here we have the variables
\ba\label{axis variables}
\ell_e(s_y,s_z)\q \text{with}\q e\in\{\varphi, \zeta,  \varphi\zeta\}  .
\ea

\subsection{Fourier transform}

The regular lattice of the boundary allows us to define a discrete Fourier transform. The Fourier transform will (block)-diagonalize the Hessians resulting from the Regge action, which will hugely simplify their analysis. We need to take into account the twist angles $\gamma_y$ and $\gamma_z$ in the background geometry (\ref{identifications}). In the triangulation we incorporate these twists by rotating the hyper-cylinder by the respective twist angles before gluing it to the three-torus.

 We write the twist angles $\gamma_i$, with $i\in\{y,z\}$, as 
\be
\gamma_i = \frac{2\pi}{N_\theta} \Upsilon_i   ,
\ee
so that the  $\Upsilon_i$ give the angles in lattice units. 
We then have the periodicities
\ba\label{period1}
\ell_e(s_\theta, N_y, s_z)\,=\, \ell_e( s_\theta -\Upsilon_y, 0,s_z) , \q \text{and} \q
\ell_e(s_\theta, s_y, N_z)\,=\, \ell_e( s_\theta -\Upsilon_z, s_y,0)   ,
\ea
for the fluctuation variables attached to the boundary. 

We define the Fourier transformation in $\theta$ in the usual way
\ba
\ell_e(k_\theta, s_y, s_z)&=& \frac{1}{\sqrt{N_\theta}} \sum_{s_\theta} e^{-2\pi \i \frac{k_\theta\cdot s_\theta}{N_\theta} } \, \ell_e(s_\theta, s_y, s_z ) 
\ea
so  that the periodicity relations  (\ref{period1}) are now given by
\ba
\ell_e(k_\theta, N_y, s_z) \,=\, e^{- \i\gamma_1  k_\theta} \ell_e(k_\theta, 0, s_z) , \q \text{and} \q
\ell_e(k_\theta, s_y, N_z)\,=\, e^{-\i\gamma_2  k_\theta} \ell_e(k_\theta, s_y , 0)  .
\ea
Thus, the phase shifted variables 
\ba
\ell^{\rm ps}_e(k_\theta,s_y,s_z) \,:=\, e^{\i \gamma_1 k_\theta \frac{s_y}{N_y}} \,e^{\i \gamma_2 k_\theta \frac{s_z}{N_z}} \, \ell_e(k_\theta,s_y,s_z)
\ea
are periodic in the usual way:
\ba
\ell^{\rm ps}_e(k_\theta, N_y,s_z) \,=\, \ell^{\rm ps}_e(k_\theta, 0, s_z)  ,\q \text{and} \q
\ell^{\rm ps}_e(k_\theta, s_y,N_z) \,=\, \ell^{\rm ps}_e(k_\theta, s_y, 0) . 
\ea
We define a 'twisted' Fourier transform in all three variables as
\ba\label{twistedFT}
\ell_e(k_\theta, k_y,k_z)&=&  \frac{1}{\sqrt{N_\theta N_y N_z }} \sum_{s_\theta,s_y,s_z}  e^{-2\pi \i (   \frac{k_\theta\cdot s_\theta}{N_\theta} + \frac{v_y\cdot s_y}{N_y}+\frac{v_z\cdot s_z}{N_z})}\, \ell_e(s_\theta, s_y,s_z),
\ea
where
\ba
v_y\,=\,k_y-\frac{\gamma_1}{2\pi} k_\theta  ,\q \text{and} \q
v_z\,=\,k_z-\frac{\gamma_2}{2\pi} k_\theta  .
\ea
For later use we introduce the abbreviations
\ba \label{omegas}
\omega_\theta\,=\, e^{  \frac{2\pi \i   k_\theta}{N_\theta} } \; ,\q \omega_y\,=\, e^{  \frac{2\pi \i   v_y}{N_y} } \; , \q  \text{and} \q \omega_z\,=\, e^{  \frac{2\pi \i   v_z}{N_z} }  .
\ea

The variables attached to the edges living in the two-dimensional axis $\ell_i(s_y,s_z)$ depend only on $s_y$ and $s_z$.  On the axis the twists have a trivial action and we can therefore just employ the usual Fourier transform
\ba\label{axisFT}
\ell_e( k_\theta,k_y,k_z)&=&  \frac{1}{\sqrt{N_\theta N_y N_z }} \sum_{s_\theta,s_y,s_z}  e^{-2\pi \i (   \frac{k_\theta\cdot s_\theta}{N_\theta}+  \frac{k_y\cdot s_y}{N_y}+\frac{k_z\cdot s_z}{N_z})}\, \ell_e( s_y,s_z) .
\ea
This is consistent with (\ref{twistedFT}): as the variables on the axis have no $\theta$ dependence, the sum over $s_\theta$ leads to an $N_\theta\,\delta_{k_\theta, 0}$-factor, so that we can set $v_y=k_y$ and $v_z=k_z$.

\section{Zeroth and  first order boundary effective actions}\label{ZeroFirst}

Having fixed the triangulation and its background geometry we can now evaluate the Regge action on solutions to the equations of motion. As the Regge equations are highly non-linear, we consider an expansion around the chosen background and evaluate the action up to second order in perturbations.

To this end we split the length variables into $l_e = L_e + \ell_e$ and expand the Regge action into zeroth, first, and second order effective actions, 
$
S_R=S^{(0)}_R +S^{(1)}_R+S^{(2)}_R +{\cal O}(\ell^3)
$. When evaluated on a solution we refer to the Hamilton-Jacobi functional and its various orders: $S_{R}{}_{|\text{sol}} := S_{HJ} = S^{(0)}_{HJ} +S^{(1)}_{HJ}+S^{(2)}_{HJ} +{\cal O}(\ell^3)$.
\\[1mm]

For the zeroth order we have to evaluate the Regge action on the flat background and therefore only need to consider the boundary term
\be
S^{(0)}_{R}{}_{|\text{sol}} \,=\, -\frac{1}{8\pi G}\sum_{t \in \partial \cal{T}} A_t(L_e) \,\psi_t(L_e) , \q \text{with } \q   \psi_t = \pi - \sum_{\sigma \supset t} \theta_t^\sigma   .
\ee
For the background triangulation we have chosen, the triangles with non-vanishing boundary deficit angles are those that lie on the two-dimensional rectangular faces with side lengths $Y$ and $Z$.  
These rectangular faces  are made up of two identical triangles both with area $\frac{1}{2}YZ$. The deficit angle associated to these triangles is also the same and given by 
$\psi_t=\frac{2\pi}{N_\theta}$. There are $2 N_\theta \cdot N_y \cdot N_z$ such triangles and  so we obtain
\ba
S^{(0)}_{R}{}_{|\text{sol}} \,=\, -\frac{1}{8\pi G} N_\theta N_y N_z \, Y Z \frac{ 2\pi}{N_\theta} \,=\,  -\frac{1}{4 G}    \alpha \beta.
\ea
This gives the zeroth order of the Hamilton-Jacobi functional $S_{HJ}^{(0)}$ both for our discretization and in the continuum. 
\\[1mm]

The first order variation of the Regge action is given by 
\be
\delta S_R \, =\,  -\frac{1}{8\pi G} \left[ \sum_{t \in  \cal{T}^\circ} \left( \frac{\partial A_t}{\partial l_e} \,\epsilon_t \right) \ell_e +  \sum_{t \in \partial \cal{T}}  \sum_{e\subset t}\left( \frac{\partial A_t}{\partial l_e} \,\psi_t \right) \ell_e \right] , 
\ee
where, as in \ref{Section:Regge}, $\cal{T}^\circ$ and $\partial \cal{T}$ refer to the bulk and boundary portions of the triangulation $\cal{T}$. On flat solutions $\epsilon_t =0$, and the bulk part vanishes. As before, only those boundary triangles with non-vanishing extrinsic curvature angle $\psi_t$ contribute, and once again these are the triangles in the rectangular faces with side lengths $Y$ and $Z$. Therefore, we get  
\ba\label{HJ1O}
S^{(1)}_{HJ}&=& -\frac{1}{8  G\, N_\theta}  \sum_{s_\theta,s_y,s_z}  \left\{ Z [\ell_y (s_\theta,s_y,s_z) + \ell_y (s_\theta,s_y\!+\!1,s_z) ]+ Y [\ell_z (s_\theta,s_y,s_z) +  \ell_z (s_\theta,s_y,s_z \! +\! 1)] \right\} \nn\\
&=& -\frac{1}{4  G\, N_\theta}  \sum_{s_\theta,s_y,s_z}  \left\{ Z \ell_y (s_\theta,s_y,s_z) + Y \ell_z (s_\theta,s_y,s_z) \right\} \nn\\
&=&  -\frac{\sqrt{N_\theta N_y N_z}}{4  G\, N_\theta} \left\{ Z \ell_y(k_\theta=0,k_y=0,k_z=0) \,+\, Y \ell_z (k_\theta=0,k_y=0,k_z=0)\right\}  . 
\ea

In section \ref{Section:Regge} we discussed the fact that there is a notion of (residual) diffeomorphisms for Regge configurations on a flat background. These diffeomorphisms act by displacing a vertex in the embedding flat space time. The vertex displacement induces a change of lengths for the edges adjacent to this vertex.  Here we are interested in describing these displacement induced length changes to first order in the fluctuation variables $\ell_e$ explicitly. A vertex can be displaced in four directions, which we can identify to be the directions (in the background geometry) of the radial edges, the edges in the $\theta$-direction and in the $y$- and $z$-directions. A displacement in an orthogonal direction (with respect to the background geometry) to a given edge will not affect the length of this edge to first order. For displacements in the radial and angular directions we therefore have $\ell_y=\ell_z=0$. For displacements in the (negative) $y$-direction by an amount $\chi_y$ and for displacements in the (negative) $z$-direction by an amount $\chi_z$ we obtain
\ba
\ell_y (s_\theta,s_y,s_z)  \,=\, \chi_y (s_\theta,s_y,s_z)  -\chi_y (s_\theta,s_y\!+\!1,s_z) ,\\
 \text{\&} \q \ell_z(s_\theta,s_y,s_z) \, = \,\chi_z (s_\theta,s_y,s_z)  -  \chi_z (s_\theta,s_y,s_z\!+\!1)\, .
\ea
As the first order of the Hamilton-Jacobi function (\ref{HJ1O}) is a sum over these edge lengths, and because of the periodicity of the $y$- and $z$-directions, we have that  $S^{(1)}_{HJ}$ vanishes for boundary perturbations that describe (boundary) vertex displacements, that is, for flat solutions (of the linearized equations of motions).  Similarly, one finds in the continuum that the first order of the Hamilton-Jacobi function is a total divergence if evaluated on boundary data describing a linearized diffeomorphisms \cite{BonzomDittrich}.  \\[1mm]

In (\ref{HJ1O}) we have used length variables to express the first order of the Hamilton-Jacobi functional. For the continuum limit it is useful to transform the length variables on the boundary to metric variables. This transformation is non-linear and this will lead to a second order contribution to the Hamilton-Jacobi functional in metric variables, coming from the first order in length variables.

Using the transformation of length variables to metric variables defined in Appendix \ref{app:lengthmetric}, we can express the length variables appearing in $S^{(1)}_{HJ}$  up to second order in metric variables as 
\be
\ell_y = \frac{1}{2Y} h_{yy} -  \frac{1}{8Y^3} (h_{yy})^2  + {\cal O} ((h_{yy})^3), \q \text{and} \q \ell_z = \frac{1}{2Z} h_{zz} -  \frac{1}{8Z^3} (h_{zz})^2  + {\cal O} ((h_{zz})^3) .
\ee 
Introducing rescaled metric variables $h'_{aq}=\frac{1}{H_{aa}} h_{aa}$ and the shorthand $N \equiv N_\theta N_y N_z$, the part of the Hamilton-Jacobi action that is first order in length variables gives the following contributions to first and second order in (rescaled) metric variables: 
\ba\label{HJFO}
S^{(1)}_{HJ} &\rightarrow& -\frac{\sqrt{N}}{ 8G N_\theta} YZ  \left( h'_{yy} + h'_{zz}\right)_{|k=0}
+\frac{YZ}{32\, G N_\theta} \!\!\sum_{k_\theta,k_y,k_z}\!\!\! \left( h'_{yy}(k) h'_{yy}(-k) + h'_{zz}(k) h'_{zz}(-k) \right)
\nn\\
&&= -\frac{1}{ \sqrt{N}}\frac{\alpha \beta}{8 G} \left( h'_{yy} + h'_{zz}\right)_{|k=0}
+\frac{YZ}{32\, G N_\theta}\!\! \sum_{k_\theta,k_y,k_z} \!\!\!\left( h'_{yy}(k) h'_{yy}(-k) + h'_{zz}(k) h'_{zz}(-k) \right), \;\;\;\;\;\;\;\;
\ea
where in the latter equality we have used $\alpha=N_y Y$ and $\beta=N_z Z$.

\section{The second order of the Regge action}\label{SecondOutline}

The second order of the Regge action (in length perturbations) is given by
\be\label{2nd1}
S_R^{(2)} \,=\, - S^{ {\cal T}^\circ} \,- S^{\partial \cal T}    \,+\,  \sum_\sigma S^\sigma 
\ee
where
\ba
& S^{ {\cal T}^\circ} =   \sum_{t\in {\cal T}^\circ} \sum_{e,e'} \ell_e\, \left(\frac{\partial^2 A_t}{\partial l_e \partial l_{e'}}\epsilon_t \right) \,  \ell_{e'},  \q  S^{\partial \cal T} = \sum_{t\in \partial {\cal T} } \sum_{e,e'} \ell_e\, \left(\frac{\partial^2 A_t}{\partial l_e \partial l_{e'}}\psi_t \right) \,  \ell_{e'} ,\q
 \ea
and 
\be
  S^\sigma = \frac{1}{16\pi G} \sum_t \sum_{e,e'} \ell_e  \left. \left(    \sum_{t \subset \sigma} \frac{\partial A_t}{\partial l_e} \frac{\partial \theta^\sigma_t}{\partial l_{e'}} \right) \right|_{l_e = L_e} \ell_{e'}  .
\ee

The first, bulk contribution $S^{ {\cal T}^\circ}$ once again vanishes because we work on a flat background where $\epsilon_t=0$.  The second term in (\ref{2nd1}) is a boundary term, whose computation is very similar to the first order term in (\ref{HJ1O}). The only non-vanishing angles are $\psi_t=\frac{2\pi}{N_\theta}$ for the boundary triangles in the $yz$-plane, and so we only have  to consider the second derivatives of the areas for these triangles. After a Fourier transform, and using the rescaled length variables $\hat{\ell}_e = L_e \ell_e$, which turns out to simplify the matrix entries, we have    
%
\be\scalemath{0.95}{
S^{\partial \cal T} \,=\,   \frac{ 1}{ 8 G N_\theta} \frac{ 1}{YZ}   \sum_{k_\theta,k_y,k_z} 
\left( \begin{array}{c}  \hat{\ell}_{y}(k)\\  \hat{\ell}_{z}(k) \\  \hat{\ell}_{yz}(k) \end{array}\right)^{\rm t} 
\left( \begin{array}{ccc} -1  & 0  & \frac{1}{2} (1+\omega_z) \\ 0  & -1  & \frac{1}{2}(1+\omega_y)  \\ \frac{1}{2}(1+\omega_z^{-1}) & \frac{1}{2}(1+\omega_y^{-1}) & -1  \end{array} \right)  
\left( \begin{array}{c}  \hat{\ell}_{y}(-k)\\  \hat{\ell}_{z}(-k) \\  \hat{\ell}_{yz}(-k) \end{array}\right)  }. \label{BDTERM1}
\ee
~\\

It is much more laborious to determine the last contribution to the second order of the Regge action in (\ref{2nd1}), which is a sum over contributions $S^\sigma$ for each four-simplex $\sigma$. To this end one has to compute the Hessian matrix 
\ba
H^{\cal T}_{ee'}\,=\, \sum_{\sigma \subset {\cal T}} \left. \left( \sum_{t \subset \sigma}  \frac{\partial A_t}{\partial l_e} \frac{\partial \theta^\sigma_t}{\partial l_{e'}} \right) \right|_{l_e = L_e} \,=\,\sum_{\sigma \subset {\cal T}} \left. H^\sigma_{ee'} \right|_{l_e = L_e}  .
\ea
We broke this calculation into the following steps:

\begin{itemize}
\item Evaluate the Hessian matrices  $H^\sigma_{ee'}$ on the geometry of each four-simplex in one hyper-prism.
\item Add these Hessian matrices to obtain the Hessian associated to the hyper-prism. To this end, length variables that define the same variable in the hyper-prism were identified with each other.
\item  `Glue' the Hessians of the hyper-prisms to get the Hessian of the full triangulation. The Fourier transformation (\ref{twistedFT}) block diagonalizes the Hessian for the full triangulation. This allows us to consider the blocks labeled by the momenta $(k_\theta,k_y,k_z)$. For $k_\theta>0$ each block $H^{\cal T}_{ee'}(k_\theta,k_y,k_z)$ is a $(4+7)\times(4+7)$ matrix, with $e\in\{r,ry,rz,ryz\}$ labeling the bulk edges and $e\in \{\theta,y,z,\theta y,\theta z,\theta y z\}$ labeling the boundary edges. For $k_\theta=0$ we have three additional rows and columns due to the edges on the two-dimensional inner axis, labelled by $e\in \{\varphi, \zeta, \varphi\zeta\}$. 
\item Compute the effective actions by integrating out all bulk edges except for the radial edges with $e=r$. The latter are not integrated out because we want to understand the variables $\ell_r$ associated to these edges as boundary fields.  This integration process starts with the variables $e\in \{\varphi, \zeta, \varphi\zeta\}$ on the two-dimensional axis. There are two gauge modes at $k_\theta=0$, which arise from the vertex displacement symmetry of the bulk vertices along this two-dimensional axis. However, this does not matter for evaluating the action on the solutions---by definition the value of the action is constant along the gauge orbit. Similarly, we will have one gauge mode for $k_\theta=1$ and for $k_\theta=-1$ arising from the vertex displacement symmetry of the bulk vertices in the directions orthogonal to the two-dimensional axis. We will deal in more detail with these gauge symmetries in section \ref{sec:oneloop}, where we compute the path integral to one-loop order.  After integrating out the axis variables we proceeded to integrate out the diagonals and hyper-diagonals in the bulk, the edges with $e\in\{ry,rz,ryz\}$.  Finally, to compute the second order of the Hamilton-Jacobi functional we integrate out the radial variables, see section \ref{section:HJF}. 
\end{itemize}

The Hessian matrices appearing at the different stages of the integration process exhibit a particular scaling in the background variables. We use this scaling to redefine our variables and simplify the matrices. Furthermore, for the Hamilton-Jacobi action, as well as for the effective action for the `radial' boundary field, we transform the length variables to metric variables. This again simplifies the expressions and the (interpretation of the) continuum limit.  

\section{Computation of the second order boundary effective action}\label{ComputeSecondEffective}

The most time consuming part of the work is the computation of the second order of the Regge action and the second order of the boundary effective action. To deal with the very lengthy expressions that can appear at intermediate stages, and will therefore not be displayed here, we have used Mathematica. This section collects a number of definitions and elaborations that will allow us to explain and interpret the results of these computations.

To begin with it is convenient to introduce various variable re-scalings in section \ref{scalings}.  We will also transform the boundary length variables to boundary metric variables, which will simplify the continuum limit.

To ease the interpretation of the results we will introduce a basis of geometrically motivated modes for the boundary metric in subsection \ref{sec:projections}. This allows us to project onto the flat sector, that is, the boundary metrics that induce a flat solution, and to define the mode describing the boundary graviton. To find these geometrically motivated modes we  have to identify  how diffeomorphisms act on the bulk and boundary variables, which we do in subsections \ref{bulkd} and \ref{boundaryd}. Finally, subsection \ref{contlimit} explains how to obtain the continuum limit from the discrete expressions.

\subsection{Variable transformations and scalings}\label{scalings}

The Hessian  $H^{\rm pr}$ for the hyper-prism simplifies if we introduce the rescaled variables 
\ba
\hat \ell_e \,=\, L_e\, \ell_e
\ea
and extract a pre-factor: we define
\ba
H^{\rm pr} \,=\,      \frac{1}{24 V_\sigma}  \text{diag}(\{L_e\}_e) \cdot M^{\rm pr} \cdot \text{diag}(\{L_e\}_e)   ,
\ea
here $V_\sigma = \frac{\Theta R Y Z}{24} \sqrt{1 - \frac{x}{2}}$, with $x\equiv \frac{\Theta^2}{2R^2} = 1-\cos \frac{2\pi}{N_\theta}$, is the four-volume of a four-simplex in the triangulation. (The volumes are the same for all types of four-simplices.) 

After having integrated out all bulk variables except for the radial ones we transform the length variables on the boundary to metric variables. There are seven length variables per boundary vertex, but only six metric variables.  The additional length variable is given by the length fluctuation of the hyper-diagonal of the cuboids.  This redundancy is dealt with by finding  a transformation that completely decouples the lengths of the hyper-diagonals from the remaining variables. Interestingly, this transformation is the same as for the computation of the 3D Regge action on a cuboid lattice \cite{Williams3D, DittFreidSpez} and given by
\ba
\scalemath{0.8}{
\begin{pmatrix} 
h_{\theta \theta} \\ h_{yy}\\ h_{zz} \\ h_{\theta y} \\h_{\theta z} \\h_{y z} \\ h_{\theta y z} \end{pmatrix} =
 \begin{pmatrix} 2 & 0 & 0 & 0 & 0 &0 &0  \\ 0 & 2 & 0 & 0 & 0 &0 &0  \\ 0 & 0 & 2 & 0 & 0 &0 &0  \\ -1 & -1 & 0 & 1 & 0 &0 &0  \\ -1 & 0 & -1 & 0 & 1 &0 &0  \\ 0 & -1 & -1 & 0 & 0 & 1 &0 \\ \frac{1}{2}(\omega_y + \omega_z) & \frac{1}{2}(\omega_\theta + \omega_z) & \frac{1}{2}(\omega_\theta + \omega_y) & -\frac{1}{2}(1 + \omega_z) & -\frac{1}{2}(1 + \omega_y) &- \frac{1}{2}(1 + \omega_\theta) & 1 \end{pmatrix} 
 \cdot
\begin{pmatrix} \hat \ell_\theta \\ \hat \ell_y \\ \hat \ell_z \\ \hat \ell_{\theta y}   \\ \hat \ell_{\theta y} \\ \hat \ell_{y z} \\ \hat \ell_{\theta y z} \end{pmatrix}
} .
\ea
In appendix \ref{app:lengthmetric} we explain in more detail the transformation from length to metric variables.  Finally we apply a further rescaling to the radial and the boundary metric variables:
\ba \label{rescaled-metric}
h'_{ab} \,=\, \frac{1}{\sqrt{H_{aa}H_{bb}}} h_{ab} \; , \q  h'_{\theta y z}=  \frac{1}{\Theta Y Z} h_{\theta y z} \;,  \q \text{and} \q \ell_r'\,=\,  \frac{1}{\Theta^2}  \hat \ell_r \,=\,  \frac{R^2}{\Theta^2}  \ell_r \,=\, \frac{1}{2x} \ell_r  ,
\ea
where $H_{ab}=\text{diag}(\Theta^2,Y^2,Z^2)$ is the boundary background metric. 

We then express the effective action for the radial boundary field as  
\ba\label{r-eff}
S^{\rm eff}_r \,=\,  \frac{1}{16\pi G} \frac{  V^2_{\rm cube}  }{24 V_\sigma}   \sum_{k_\theta,k_y,k_z} 
\left( \begin{array}{c}  \ell'_r(k)\\ h' (k) \\  h'_{\theta x y}(k) \end{array}\right)^{\rm t} \circ
\left( \begin{array}{ccc} M_{rr}(k)   & ({B}(k))^{\rm t}  & 0 \\ { B}(-k) & M_{h}(k) & 0 \\ 0 & 0 & M_{\theta y z}(k) \end{array} \right)  
\circ
\left( \begin{array}{c} \ell'_r(-k) \\ h'(-k) \\ h'_{\theta y z}(-k) \end{array}   \right) 
\ea
where $(h')^{\rm t}=(h'_{\theta\theta}, h'_{yy}, h'_{zz}, h'_{\theta y}, h'_{\theta z}, h'_{yz})$ summarizes the boundary metric variables. We use the circle product $\circ$ to denote the multiplication of boundary metric vectors $f'$ and $g'$ such that
\ba\label{circconv}
(f')^{\rm t} \circ g' \,:=\,  \sum_{a} f'_{aa} g'_{aa}  \,+\, 2 \sum_{b>a} f'_{ab} g'_{ab}\, ,
\ea
 where $a$ and $b$ take values in $\{\theta,y,z\}$ and are ordered according to $\theta<y<z$. This convention reproduces the usual inner product for metric fluctuations. 

The three-volume of a cuboid in the boundary lattice is $V_{\rm cube}=\Theta Y Z$, and so the pre-factor is
\ba\label{r-eff2}
 \frac{  V^2_{\rm cube}  }{24 V_\sigma}  \,=\, \frac{\Theta Y Z} { R \sqrt{1 - \frac{x}{2}} }\,=\, \frac{\Theta Y Z} { R \sqrt{1 - \frac{\Theta^2}{4R^2}} }    .
\ea
Finally, we introduce short hands for various (rescaled) difference operators, which will appear in $M_{rr}$ and $M_h$: 
\begin{alignat}{6} \label{d-def}
 \omega_\theta \,&=&\, e^{  \frac{2\pi \i   k_\theta}{N_\theta} } \; , \q \omega_y \, &=&\, e^{  \frac{2\pi \i   v_y}{N_y} } \; , \q  \omega_z \,&=&\, e^{  \frac{2\pi \i   v_z}{N_z} }\; , \nn\\
 d_\theta \,&=&\, \frac{1-\omega_\theta}{\Theta} \; , \q  d_y \, &=&\, \frac{1-\omega_y}{Y} \; , \q  d_z \,&=&\, \frac{1-\omega_z}{Z} \; , \nn\\
 d^*_\theta \,&=&\, \frac{1-\omega^{-1}_\theta}{\Theta} \; , \q  d^*_y \, &=&\, \frac{1-\omega^{-1}_y}{Y} \; , \q  d^*_z \,&=&\, \frac{1-\omega^{-1}_z}{Z} \; , \nn\\
 \Delta_\theta\, &=&\, d_\theta d^*_\theta \; , \q  \Delta_y\, &=&\, d_y d^*_y \; , \q  \Delta_z\, &=&\, d_z d^*_z \;  .
\end{alignat}

\subsection{Bulk diffeomorphisms}\label{bulkd}

As explained in section \ref{Section:Regge}, linearized Regge calculus on a flat background exhibits gauge symmetries that are discrete remnants of diffeomorphism symmetry. Indeed, in \cite{Williams1,Williams2} it is shown that the null modes of the quadratic action for linearized Regge calculus on a regular cubic lattice represent a discretization of the spin 1 modes of the metric degrees of freedom. 

To identify these null modes one considers a displacement of the vertex in the embedding flat geometry and computes the induced change of the length variables of the adjacent edges to first order in the displacement parameter (e.g. the lengths of the displacement in the background geometry). See \cite{BonzomDittrich} for explicit computations in the 3D context, which motivated the 4D example considered here. 

The gauge degrees of freedom are associated to the bulk vertices. In our example we only have bulk vertices on the two-dimensional axis. The vertices of this axis can be displaced in the radial, the angular, and the $y$ and $z$ directions. The last two displacements only effect the $k_\theta=0$ modes of the various bulk variables and are given by 
\ba\label{bulkdiff1} 
(n^{\hat{\ell}}_y(k))^{\rm t} = \left( 0, \omega_y, 0, \omega_y \omega_z, \frac{1}{\sqrt{N_\theta}} (\omega_y -1), 0, \frac{1}{\sqrt{N_\theta}} (\omega_y \omega_z -1)  \right) \\
(n^{\hat{\ell}}_z(k))^{\rm t} = \left( 0, 0, \omega_z,  \omega_y \omega_z, 0, \frac{1}{\sqrt{N_\theta}} (\omega_z -1),  \frac{1}{\sqrt{N_\theta}} (\omega_y \omega_z-1)  \right)
\ea 
with the entries in the order $(\hat{\ell}_r, \hat{\ell}_{ry}, \hat{\ell}_{rz}, \hat{\ell}_{ryz}, \hat{\ell}_{\varphi}, \hat{\ell}_{\zeta}, \hat{\ell}_{\varphi \zeta} )$.
The radial variables are not affected as the radial edges are orthogonal to the axis. The diagonals with an $r$-component change by a $\theta$-independent amount; this is the reason why this gauge symmetry only involves the $k_\theta=0$ mode. 

Finally, there is the displacement of the bulk vertices orthogonal to the axis, that is, in the $r\theta$-plane.  This displacement will leave the axis variables unaffected. The change in the radial variables can be found by considering a central vertex in a disk connected by $N_\theta$ equally distributed edges to the boundary of the disk: We parametrize this boundary by $\theta\in [0,2\pi)$. Displacing the central vertex, e.g. along the $\theta=0$ line, induces a change in the lengths of these edges proportional to $\cos(\theta)$. For a displacement along the $\theta=\pi/2$ line we obtain a change proportional to $\sin(\theta)$. Hence this gauge symmetry involves only the $k_\theta=+1$ and $k_\theta=-1$ modes and is given by
\ba\label{bulkdiff2}
(n^{\hat{\ell}}_{\pm 1}(k))^{\rm t} = \left( 1 , \omega_y , \omega_z, \omega_y \omega_z, 0, 0, 0  \right) .
\ea

These gauge symmetries can easily be fixed and do not present a problem for computing the effective action. However, they do need to be considered more carefully for the computation of the one-loop determinant in section \ref{sec:oneloop}.

\subsection{Diffeomorphisms affecting the boundary}\label{boundaryd}

Similarly to the vertex displacements of the bulk vertices, we can consider vertex displacements of the boundary vertices.  Applying such a vertex displacement to a flat solution will not change its flatness. Thus, considering how these vertex displacements affect the boundary metric, we will identify the space of boundary conditions that lead to flat solutions.  This will allow us to split the boundary conditions into two sets: a flat sector that entails flat solutions, and a curved sector that leads to solutions with curvature, i.e. non-vanishing deficit angles.

The vertex displacements tangential to the boundary itself are described by the following vectors (in the rescaled boundary metric variables $h'_e$):
 \ba\label{bdrydiff}
(n^{h'}_\theta(k))^{\rm t}&=& (-2d^*_\theta, 0,0,  d_y,  d_z,0) \, ,\nn \\
(n^{h'}_y(k))^{\rm t} &=&(0, -2d^*_y,0, d_\theta, 0,  d_z)\, , \nn\\
\text{and} \ \ (n^{h'}_z(k))^{\rm t} &=&(0,0, -2d^*_z, 0,  d_\theta,  d_y)\, ,
\ea 
where the entries are given in the order $(h'_{\theta \theta}, h'_{yy}, h'_{zz}, h'_{\theta y}, h'_{\theta z},h'_{yz})$. 

One could expect that the quadratic part of the Hamilton-Jacobi function has these boundary diffeomorphisms as null vectors. However, this is not the case in general. The reason is that there is a non-vanishing first order term in the Hamilton-Jacobi function. To make the first order term invariant under diffeomorphisms to higher than linear order requires `compensating' terms in the second order part, and these will, in general, appear as boundary diffeomorphism violating terms. 

Below we will see that the matrix $M_h$ in \eqref{r-eff} as well as  the Hamilton-Jacobi action in section \ref{section:HJF} have two parts with different scalings in $x$, and hence in the background radius $R$  of the solid three-torus. The part that dominates for large $R$ is invariant under diffeomorphisms tangential to the boundary.  In addition $M_{HJ}$ has the boundary diffeomorphisms in $y$ and $z$ direction as null vectors.

Finally, we have vertex displacements on the boundary in the radial direction. These are described by the vector
\ba\label{radialdiff}
(n_r^{h'}(k))^{\rm t}&=& 
 \,\left(2 \tfrac{1+\omega_\theta}{\Theta \, \omega_\theta},0,0, -  d_y, - d_z,0\right) \nn\\
&=&(\tfrac{4}{\Theta},0,0,0,0,0)-(n_\theta^{h'}(k))^{\rm t}  .
\ea 

\subsection{Projections on flat and curved solutions}\label{sec:projections}

Boundary metric perturbations induce either a flat solution or a solution with curvature. The space of boundary metrics $h_{ab}$ that induces flat solutions is spanned by the vectors describing the boundary diffeomorphisms (\ref{bdrydiff}) and the radial diffeomorphisms (\ref{radialdiff}). To define an orthogonal subspace to these metrics, we specify an inner product on the space of boundary metric perturbations. The inner product is defined for each mode $(k_\theta,k_y,k_z)$ separately:
\ba
\langle  h^1 \, , h^2 \rangle (k) \,=\, \sum_{a,b} h^1_{ab} (k) \,\, \tfrac{1}{2} \left( H^{ac}H^{bd}+ H^{ad}H^{bc}\right)\,h^2_{cd} (-k) ,
\ea
where $H^{ab}=\text{diag}(\Theta^{-2},Y^{-2},Z^{-2})$ is the inverse of the boundary background metric.  With the rescaled variables 
$h'_{ab}=   \sqrt{H^{aa} H^{bb}}  h_{ab}$   we can write this inner product as
\ba
\langle  h^1 \, , h^2 \rangle \,=\ \sum_{a} h'^1_{aa} h'^2_{aa}  \,\, + 2 \, \sum_{a<b} h'^1_{ab} h'^2_{ab} \, \,=:\, (h'^1)^{\rm t} \circ h'^2  ,
\ea
using again the convention defined in (\ref{circconv}).

The boundary and radial diffeomorphisms are {\it not} orthogonal to each other, nor are the different types of boundary diffeomorphisms mutually orthogonal.  To build a projector onto the space spanned by each of these diffeomorphisms we would have to go through an orthonormalization procedure for the corresponding vectors. However, there is a short cut---we use the spin projectors for the  background geometry of the boundary.

The background boundary geometry is flat and we can simply define  the spin 0, spin 1, and spin 2 projectors. These projectors are generally useful, e.g. the quadratic action for 3D gravity on a flat background can be written as a sum of the spin 0 and spin 2 projectors. This is possible because of the rotational symmetry of the background. Here, although the background boundary metric has the same symmetry, its embedding into the 4D spacetime breaks the symmetry and, indeed, the boundary effective action will not be a sum of spin projectors. 

Nevertheless, the projectors are quite useful: the spin 1 projector determines the space of diffeomorphisms tangential to the boundary. We use this projector to construct the  diffeomorphism component in the radial direction, which is orthogonal to the tangential boundary diffeomorphisms.  This allows us to construct a projector onto the orthogonal part of the radial diffeomorphisms. The remaining vector space of dimension two is spanned by boundary fluctuations inducing curved solutions.

These projectors can also be defined on a lattice of rectangular cuboids \cite{BahrDittrichHe}, which we are using for the boundary discretization. In this context, the spin 1 projector describes the discrete boundary diffeomorphisms, which are also symmetries of the linearized three-dimensional Regge action \cite{Williams3D, DittFreidSpez}. 

The discrete projectors on the space of rescaled boundary fluctuations $h'_{ab}$ are given by  
\ba\label{DiscretePr}
\Pi^{(0)}_{ab\, cd}  &=& 
\frac{1}{2} \left(
 \delta_{ab} + (1-\delta_{ab})\frac{ d^*_a d^*_b}{\Delta} -\delta_{ab} \frac{ d_a d^*_b}{\Delta}
\right)  \left(
 \delta_{cd} + (1-\delta_{cd})\frac{ d_c d_d}{\Delta} -\delta_{cd} \frac{ d^*_c d_d}{\Delta}
 \right),
 \nn\\
\Pi^{(2)}_{ab\, cd}  &=& 
\frac{1}{2}( (1-\delta_{ab}) +\delta_{ab} \omega_b   )  ((1-\delta_{cd}) +\delta_{cd} \omega^{-1}_d) \times \nn\\
&&
\left( \left( \delta_{ac}-\frac{d^*_ad_c}{\Delta}\right)\left(\delta_{bd} -\frac{d^*_bd_d}{\Delta} \right)+
\left( \delta_{ad}-\frac{d^*_ad_d}{\Delta}\right)\left(\delta_{bc} -\frac{d^*_bd_c}{\Delta} \right)
\right)
- \Pi^{(0)}_{ab\, cd} , \nn\\
\text{and} \ \  \Pi^{(1)}_{ab\, cd}  &=& {\mathbb I}_{ab\, cd}   \,-\,  \Pi^{(2)}_{ab\, cd}  \,-\, \Pi^{(0)}_{ab\, cd}  ,
\ea
where ${\mathbb I}_{ab\, cd} =\tfrac{1}{2}(\delta_{ac}\delta_{bd}+\delta_{ad}\delta_{bc})$ is the identity on the space of symmetric tensors of rank two.  We have abbreviated $\Delta=\Delta_\theta+\Delta_y+\Delta_z$.

The image of the spin 1 projector is spanned by vectors
\ba
v^c_{a b}(k)\,=\, \delta^c_a d_b ( \delta_{ab}+(1-\delta_{ab}) \omega_a) \,+\, \delta^c_b d_a( \delta_{ab}+(1-\delta_{ab} \omega_b) ) ,
\ea
which, modulo a phase, agree with the diffemorphisms tangential to the boundary described in section \ref{boundaryd}.

These discrete projectors satisfy the usual requirements for orthogonal projectors, that is $\Pi^{(i)} \circ \Pi^{(j)}= \delta^{ij} \Pi^{(j)}$. As discussed above we can use these projectors to construct a vector $V^\perp$. This (normalized) vector describes a metric perturbation leading to a flat solution, but  is orthogonal to the fluctuations induced by diffeomorphisms tangential to the boundary. Therefore $V^\perp$ can be identified with the boundary graviton mode. 

With $V^\perp$ in hand, we can construct the projector $\Pi^{\text{curv}}$ onto the space of boundary metric fluctuations that induce curved solutions:  
\ba\label{onbasis1}
(V^\perp(k))^{\rm t} &=&  \frac{1}{\Delta} 
\left( 
(\Delta_y+\Delta_z),\, \frac{\Delta_y \Delta_\theta}{\Delta_y+\Delta_z} ,\, 
\frac{\Delta_z \Delta_\theta}{\Delta_y+\Delta_z},\,
 d_y d_\theta,\,  d_z d_\theta, \, - \frac{\Delta_\theta d_y d_z}{\Delta_y+\Delta_z}
 \right) \, ,\nn\\
\Pi^{\perp}_{ab \, cd}&=&  V^\perp(-k)_{ab}  \, V^\perp(k)_{cd} \, ,\nn\\
\text{and} \ \ \Pi^{\text{curv}}_{ab\, cd}&=& \Pi^{(0)}_{ab\, cd}+ \Pi^{(2)}_{ab\, cd} \,-\, \Pi^{\perp}_{ab\, cd}  .
\ea
In fact, $\Pi^{\text{curv}}$ does project on a two-dimensional subspace of boundary conditions inducing solutions with curvature. According to the last line in (\ref{radialdiff}), which shows that   the space of flat solutions includes vectors of the form $v_{ab}=\delta^\theta_a \delta^\theta_b$, the projector $\Pi^{\text{curv}}$ has vanishing entries in the $\theta\theta$-row and $\theta\theta$-column.  The curved sector, then,  is spanned by the following (orthonormalized) basis:
\ba\label{onbasis2}
(W^{\text{curv}}(k))^{\rm t} &=& \frac{1}{\Delta_y+\Delta_z} (0, \Delta_z, \Delta_y, 0,0, d_y d_z) , \ \ \text{and} \ \ \nn\\
(X^{\text{curv}}(k))^{\rm t} &=&  \frac{1}{\sqrt{2\Delta}(\Delta_y+\Delta_z)}(0,2 d^*_\theta d^*_y d^*_z, -2 d^*_\theta d^*_y d^*_z, d^*_z (\Delta_y+\Delta_z),-d^*_y(\Delta_y+\Delta_z), d^*_\theta(\Delta_y-\Delta_z)).\nn\\
\ea

We will also need the part of the angular diffeomorphisms that is orthogonal to the diffeomorphisms in the $y$- and $z$-directions. The corresponding normalized vector is
\ba\label{onbasis3}
(U^{\text{adiff}}(k))^{\rm t}&=& \frac{\sqrt{2\Delta-\Delta_\theta}}{2 \Delta} \bigg(  (n^{h'}_\theta(k))^{\rm t} 
\; -\frac{d^*_\theta d_y}{{ 2\Delta-\Delta_\theta }}(n^{h'}_y(k))^{\rm t}-\frac{d^*_\theta d_z}{{2\Delta-\Delta_\theta }}(n^{h'}_z(k))^{\rm t}\bigg)\, .\q\; 
\ea

\subsection{Continuum limit}\label{contlimit}

Our chosen triangulation is well-adapted to taking a continuum limit on the boundary. Define  edge lengths by
\[  \Theta = \varepsilon \Theta_0, \q Y = \varepsilon Y_0, \q \text{and} \q Z = \varepsilon Z_0,  \]
and take the limit $\epsilon \rightarrow 0$ while increasing $N_\theta, N_y$, and $N_z$, so that $2\pi R$, $\alpha$, and $\beta$ stay constant. Then we have
\ba
\frac{2\pi}{N_\theta} = \arccos(1-x) =  \frac{\Theta_0}{R} \varepsilon +  {\cal O}(\varepsilon^3) , \q \frac{2\pi}{N_y}= \frac{2\pi Y_0}{\alpha} \varepsilon, \q  \& \q \frac{2\pi}{N_z}= \frac{2\pi Z_0}{\beta} \varepsilon  , 
\ea
and thus
\ba\label{dCon}
d_\theta\,=\, - \frac{2\pi \i} {2\pi R} k_\theta + {\cal O}(\varepsilon) \,, \q d_y\,=\, - \frac{ 2\pi \i}{\alpha} v_y + {\cal O}(\varepsilon)\, ,\q \& \q d_z\,=\, - \frac{ 2\pi \i}{\beta} v_z + {\cal O}(\varepsilon) .
\ea
Taken as operators, the  $d_a$ are diagonal in the Fourier basis and (\ref{dCon}) gives the eigenvalues for these operators. To match the $d_a$ to differential operators in the continuum, we introduce coordinates  (see (\ref{contmetric}) for our original continuum metric and coordinates)
\ba
t_\theta \,=\, \frac{R}{\Theta_0} \theta \, ,\q t_y\,=\, \frac{1}{Y_0} y \, ,\q \& \q t_z=\frac{1}{Z_0} z\, ,
\ea
so that $t_\theta \in [0, 2\pi R/\Theta_0), t_y \in [0, \alpha/Y_0)$ and $t_z \in [0, \beta/Z_0)$. The continuum background (boundary) metric is then $H^{\rm cont}_{ab}=\text{diag}(\Theta_0^2,Y^2_0,Z^2_0)$.  We define the continuum Fourier transform as
\ba
f(k_\theta,k_y,k_z)\,=\, \sqrt{\frac{\Theta_0 Y_0Z_0}{ 2\pi R \alpha \beta} }\int dt_\theta dt_y dt_z  f(t_\theta,t_y,t_z)  e^{-2\pi \i  \left(\tfrac{ \Theta_0}{2\pi R}k_\theta t_\theta  + \tfrac{Y_0}{\alpha} v_y t_y + \tfrac{Z_0}{\beta} v_z t_z\right) }, 
\ea
with $k_a \in \mathbb{Z}$. To have a matching spectrum, at least  for $k_a << N_a$, between the discrete and continuum operators, we have to identify: 
\ba\label{difftodiff}
d_a  \, \rightarrow  -\frac{1}{\sqrt{H^{\rm cont}_{aa}}} \partial_a  \, , \q d^*_a  \, \rightarrow  \frac{1}{\sqrt{H^{\rm cont}_{aa}}} \partial_a   \, ,                       \q  \text{and} \q \Delta_a \, \rightarrow  -  \frac{1}{H^{\rm cont}_{aa}} \partial_a \partial_a  .
\ea

Now the difference operators $d_a,d^*_a$, and $\Delta_a$ we have introduced, have a straightforward translation into the continuum theory. We will see that---apart from global pre-factors---the only remaining $\varepsilon$-dependent quantity that we will encounter in $M_h$ is $\Theta=\Theta_0\varepsilon$.  These  terms,  with $\Theta$ or $\Theta^2$ factors (and no accompanying $1/x \sim 1/\Theta^2$), will  vanish in the continuum limit.

Another length variable that will appear explicitly in the Hamilton-Jacobi action is the radius $R$, it appears via $\tfrac{\Theta^2}{2x}=R^2$. The Hamilton-Jacobi action will have terms that either scale with $R^{+1}$ or with $R^{-1}$, and we will be most interested in the terms with the dominant $R$ scaling.


\section{Effective action for the radial  field with $|k_\theta| \geq 2$}\label{Effectivek2up}

We will now detail the effective action for the radial lengths variables, as defined in (\ref{r-eff}):
\ba\label{8.1}
S^{\rm eff}_r \,=\,  \frac{1}{16\pi G} \frac{  V^2_{\rm cube}  }{24 V_\sigma}   \sum_{k_\theta,k_y,k_z} 
\left( \begin{array}{c}  \ell'_r(k)\\ h' (k) \\  h'_{\theta x y}(k) \end{array}\right)^{\rm t} \circ
\left( \begin{array}{ccc} M_{rr}(k)   & ({B}(k))^{\rm t}  & 0 \\ { B}(-k) & M_{h}(k) & 0 \\ 0 & 0 & M_{\theta y z}(k) \end{array} \right)  
\circ
\left( \begin{array}{c} \ell'_r(-k) \\ h'(-k) \\ h'_{\theta y z}(-k) \end{array}   \right)  .
\ea
Integrating out the radial length variables $\ell'_r$ we will obtain the Hamilton-Jacobi action as a functional of the boundary metric variables $h'$, including the hyper-diagonal variable $h'_{\theta y z}$.

In this subsection we will assume that $|k_\theta|\geq 2$. We will consider the cases $k_\theta =0$ and $k_\theta =\pm 1$ in the next subsection \ref{Lowmodes}. 

The matrix $M_h$, which describes the boundary-boundary couplings, splits into two terms with different scaling behaviour in $R$ ,
\ba\label{Mh}
M_h \,=\, R^2 \, M^{(R^2)}_h \,+\, M^{(R^0)}_h \,\, \,=\, \,\,  \frac{\Theta^2}{2x}M^{(R^2)}_h \,+\, M^{(R^0)}_h  .
\ea
To extract this scaling we only consider explicit dependencies on $R$ via $x=\Theta^2/(2R^2)$. There is also an indirect dependence on $R$ via $\Delta_\theta(k_\theta) =2-2\cos(2\pi k_\theta/N_\theta)$ and the fact that $N_\theta$ is fixed by the relation $x=1-\cos(2\pi/N_\theta)$.  In particular, $1/\Delta_\theta=R^2$ for $k_\theta=\pm 1$. Thus, in order to conclude that $M^{(R^2)}$ dominates in the large radius limit, we should assume that $|k_\theta|>>1$. 

We will perform  an analysis of the $1/\Delta_\theta$ terms in section \ref{Lowmodes},  and show that the $1/\Delta_\theta$ terms in $M_h$ are cancelled by matching terms arising from integrating out the radial length fluctuations. Thus, if we take all these terms together, we do not have $1/\Delta_\theta$ terms, from which an additional positive $R$ scaling can arise.  For this analysis we  assume that $\Delta_y+\Delta_z \neq 0$; we will discuss the zeros of $\Delta_y+\Delta_z$ in section \ref{sec:singular}.

It turns out that  $M^{(R^2)}$ has quite a simple structure. It is invariant under the tangential boundary diffeomorphisms, which manifests itself through the corresponding vectors being null vectors of $M^{(R^2)}$.  A left and right projection with $\Pi^\perp$ annihilates this  matrix:
\ba
  \Pi^{\perp} \, \circ \, M^{(R^2)}_h \, \circ \, \Pi^{\perp} \,=\, 0 .
\ea
This shows that the Hamilton-Jacobi action, evaluated on (linearized) flat solutions, does not have a contribution from $M^{(R^2)}_h$. Indeed, we will see that the terms with the dominant $R$ scaling arise from integrating out the radial length variables. In this sense, the radial length variables will define a `dual' boundary field, whose integration gives the dominant contribution to the Hamilton-Jacobi action for flat solutions.

With the basis vectors we introduced in (\ref{onbasis1}-\ref{onbasis3}), the matrix $M^{(R^2)}_h$ is given by
\be\label{MhR}
M^{(R^2)}_h(k) =  \frac{ \Delta}{8}\left(     X^{\rm curv}_- X^{\rm curv}_+\,-\,    V^\perp_-W^{\rm curv}_+ \,-\,    W^{\rm curv}_-V^\perp_+\right)\, , \nn\\
\ee
where the $\pm$-subindices stand for $A_-=A(-k)$ and $A_+=(A(k))^{\rm t}$. Note that 
\ba
\left(     X^{\rm curv}_- X^{\rm curv}_+\,-\,    V^\perp_- W^{\rm curv}_+ \,-\,    W^{\rm curv}_- V^\perp_+\right) \,=\, (\Pi^{(2)} -\Pi^{(0)} ) ,
\ea
where $\Pi^{(i)}$ are the discrete spin projectors introduced in (\ref{DiscretePr}).  Now $(\Pi^{(2)} -\Pi^{(0)} )$ is also the combination of projectors that appears in the second order expansion of the 3D Einstein-Hilbert action  (or the Regge action in a discretization \cite{BahrDittrichHe}) on a flat background: 
\ba
  \Theta Y Z   \sum (h'(k))^{\rm t}\circ   M^{(R^2)}_h(k)  \circ  h' (-k) \,=\,  \frac{1}{2} \left[ \int \sqrt{h} {\cal R} \, d^3x \right]^{(2)} . 
\ea
Thus, it is $M^{(R^2)}_h$ that leads to a contribution to the Hamilton-Jacobi action that  is proportional to the integrated Ricci-scalar of the boundary metric. \\[1mm]

For the sub-leading term $M^{(R^0)}_h$ we find a much more involved expression: 
\ba\label{Mh0}
M^{(R^0)}_h(k) &=&   \,\,\,\,
 \frac{\Delta}{16}  \, \left( \frac{8}{(\Delta+3\Delta_\theta)}  \,-\, \Theta^2  \right)  \,  X^{\rm curv}_- X^{\rm curv}_+ \,\,+\,\,
 \frac{1}{16} \left(1- \frac{\Delta}{\Delta _{\theta }}\right)  \,\,W^{\rm curv}_- W^{\rm curv}_+ \nn\\
&&+  \frac{\Delta}{16} \left( \Theta ^2-\frac{2}{\Delta _{\theta }} - 2 \frac{\Delta-\Delta_\theta}{\Delta^2} \omega_\theta^* \right) \, V^\perp_- W^{\rm curv}_+ \,+\,  
  \frac{\Delta}{16} \left( \Theta ^2-\frac{2}{\Delta _{\theta }}  - 2 \frac{\Delta-\Delta_\theta}{\Delta^2} \omega_\theta \right) \,  W^{\rm curv}_- V^\perp_+   \nn\\ 
  &&+ \frac{\left(\Delta -\Delta _\theta\right)}{8}  \left(\Theta ^2-\frac{2}{\Delta _{\theta }}-\frac{2}{ \Delta} +\frac{ \Delta _{\theta } }{\Delta ^2}\right) \, V^\perp_- V^\perp_+  \nn\\
  &&+  \frac{\sqrt{2\Delta - \Delta_\theta}}{8\Delta}(d^*_\theta - d_\theta)\omega_\theta  \,  U^{\text{adiff}}_- W^{\text{curv}}_+ \,+\,  \frac{\sqrt{2\Delta - \Delta_\theta}}{8\Delta}(d_\theta - d^*_\theta) \omega^*_\theta \,  W^{\text{curv}}_- U^{\text{adiff}}_+\,  \nn\\
&&+  \frac{1}{8} \frac{  \sqrt{2\Delta - \Delta_\theta}}{\Delta^2} \left( \Theta(\Delta - \Delta_\theta)^2 - {(2\Delta-\Delta_\theta)} (d_\theta - d^*_\theta)  \right) \omega_\theta \,  U^{\text{adiff}}_-V^\perp_+   \nn \\ 
&&+  \frac{1}{8} \frac{  \sqrt{2\Delta - \Delta_\theta}}{\Delta^2} \left( \Theta(\Delta - \Delta_\theta)^2 - {(2\Delta-\Delta_\theta)} (d^*_\theta - d_\theta)  \right)\omega^*_\theta  \,  V^\perp_-U^{\text{adiff}}_+  \nn\\
&& + \frac{1}{4} \left(2- \frac{3\Delta_\theta}{\Delta } - \frac{\Delta_\theta^2}{\Delta^2 }\right) \,  U^{\text{adiff}}_- U^{\text{adiff}}_+\,\, . \nn\\
\ea 
As explained in section \ref{contlimit} these expressions can be straightforwardly translated to the continuum. The difference operators $d_a,d_a^*$, and $\Delta$ translate into rescaled differential operators as detailed in (\ref{difftodiff}).  Terms with $\Theta$ or $\Theta^2$ pre-factors vanish in the continuum limit and $\omega_\theta$ and $\omega^*_\theta \rightarrow 1$. 

~\\
Next we consider the terms that give the action for the radial variables $\ell'_r$. 
The diagonal coefficient $M_{rr}$  is
\ba\label{Mrr1}
M_{rr}= {2x} \left(\Theta^2-\frac{ 2 x}{\Delta_\theta}\right) \left( \Delta_y+\Delta_z \right)   ,
 \ea
 whereas the $M_{r h}$ entries are 
 \ba
 (B(k))^{\rm t} &=& -\frac{1}{4}\left(\Theta^2-\frac{ 2 x}{\Delta_\theta}\right)
 (  2\left( \Delta_y+\Delta_z \right) \, ,\,\,  \left(2 \Delta_\theta+\Delta_z \right) \, ,\,\,   \left(2 \Delta_\theta+\Delta_y \right) \, ,\, 
 2 d_{\theta } d_y \, ,\,\, 2 d_{\theta } d_z \, ,\,\,   d_y d_z \,\,)\nn\\
 &=&  -\frac{1}{4}\left(\Theta^2-\frac{ 2 x}{\Delta_\theta}\right) \left(  2\Delta V^\perp_+ +  ( \Delta +\Delta_\theta) W^{\text{curv}}_+ \right) .
 \ea
 Note that, apart from the pre-factor,  the entries of the vector $B$ are local (i.e. do not involve inverse Laplacians), whereas the vector $V^\perp$, describing the boundary-orthogonal diffeomorphisms, is non-local, even after multiplying it with the Laplacian. The added part from the curvature sector is such that it restores the locality of the $B$-vector. 
 
Finally, the hyper-diagonal variables' contribution is described by
\ba
M_{\theta y z} &=&-\frac{1}{2} \left( \Theta^2 -\frac{2 \Theta^2}{2x} \right) .
\ea

\section{Effective action for the $k_\theta=0$ and $k_\theta=\pm 1$ modes}\label{sec:low}

The modes $k_\theta=0$ and $k_\theta=\pm 1$ are subject to gauge symmetries. The corresponding null vectors are discussed in section \ref{bulkd}.  For $k_\theta=0$ we have two null vectors, and hence we expect two gauge parameters. These gauge parameters can be chosen as $\hat\ell_\varphi$ and $\hat\ell_\zeta$ and one therefore only needs to integrate out $\hat\ell_{\varphi\zeta}$ and $(\hat\ell_{ry},\hat\ell_{rz},\hat\ell_{ryz})$ to find the radial effective action.  With the same notation as in (\ref{8.1}) we have
\ba\label{Mrr0}
M_{rr}(k_\theta=0)= 2x \, \Theta^2 \left( \frac{8x}{\Theta^2} + \Delta_y+ \Delta_z \right)  .
 \ea
 The off-diagonal term $(r-h)$ is given by
 \ba
 (B(k_\theta=0))^{\rm t} &=& -\frac{1}{4}\Theta^2
 \left(  2 \left( \frac{8x}{\Theta^2} + \Delta_y+ \Delta_z \right)  \, ,\,\,  \Delta_z  \, ,\,\,  \Delta_y \, ,\, 
0 \, ,\,\, 0 \, ,\,\,   d_y d_z \,\,\right)\nn\\
 &=&  -\frac{1}{4}\Theta^2  \left[  2 \left( \frac{8x}{\Theta^2} + \Delta_y+ \Delta_z \right) V^\perp_+   + ( \Delta_y +\Delta_z)W^{\text{curv}}_+\right] \q 
 \ea
and for $M_h$ we find
\ba
M_h(k_\theta=0) &=& \frac{\Theta^2}{2x} \frac{(\Delta_y+\Delta_z)}{8} \left(     X^{\rm curv}_- X^{\rm curv}_+ -    V^\perp_- W^{\rm curv}_+ -    W^{\rm curv}_- V^\perp_+ \right)\, \nn \\
&&-\frac{\Theta^2(\Delta_y+\Delta_z) }{16}\left(     X^{\rm curv}_-X^{\rm curv}_+ -    V^\perp_- W^{\rm curv}_+ -    W^{\rm curv}_- V^\perp_+\right)\, \nn \\
&&+ \frac{\Theta^2}{8}  \left( \frac{8}{\Theta^2} + \Delta_y+ \Delta_z \right) V^\perp_- V^\perp_+ \,+\,
 \frac{1}{2}\left(  X^{\rm curv}_- X^{\rm curv}_++  U^{\rm adiff}_-U^{\rm adiff}_+   \right) \nn \\
&&+  \frac{1}{8} \Theta \sqrt{2(\Delta_y+\Delta_z)} \left( V^\perp_- U^{\rm adiff}_+ + U^{\rm adiff}_-V^\perp_+  \right) .
\ea
Finally, we have for the hyper-diagonal contribution
\be
M_{\theta y z}(k_\theta=0) \,=\, -\frac{1}{2} \left(\Theta^2- \frac{2\Theta^2}{2x} \right) ,
\ee
as for general $k_\theta$.

For $k_\theta=+1$ and for $k_\theta=-1$ we have one null eigenvector. The gauge parameter can be identified with the radial fluctuation $\ell'_r$. Integrating out all the bulk variables, except the radial ones, we find that  $M_{rr}(k_\theta=\pm 1)$ and  $B(k_\theta=\pm 1)$ vanish. This result, as well as $M_h$ and $M_{\theta y z}$, can be found by setting $\Delta_\theta(k_\theta=\pm 1)=\tfrac{2x}{\Theta^2}$ in the formulas for the general case.

 \section{The Hamilton-Jacobi functional} \label{section:HJF}
 
Integrating out the radial variable $\ell'_r$ we arrive at the Hamilton-Jacobi action, which is a functional of the boundary metric. There are five contributions to the Hamilton-Jacobi action:
 \begin{enumerate}
 \item  The term $M_h$, detailed in (\ref{MhR}) and (\ref{Mh0}), is the boundary-boundary part of the radial effective action.  According to (\ref{r-eff}-\ref{r-eff2}) this leads to the following contribution to the Hamilton-Jacobi action
 \ba
 S_{HJa}^{(2)}&=& \frac{1}{16\pi G} \frac{\Theta Y Z}{R \sqrt{1-\frac{\Theta^2}{4R^2}}} \sum_{k} (h'(k))^{\rm t} \circ \left( R^2 M^{(R^2)}_h+M^{(R^0)}_h\right) \circ h'(-k)  .
 \ea
 \item Integrating out the radial variables, we obtain a matrix 
 \ba\label{Nhdef}
 N_h (k)&=&  -\frac{1}{M_{rr}}  B(-k)     (B(k))^{\rm t} \nn\\
 &=& -\frac{1}{16} \big(R^2 -\frac{1}{\Delta_\theta}\big) \frac{1}{\Delta-\Delta_\theta}
 \bigg( 4 \Delta^2 \,V^\perp_-V^\perp_+ +
 (\Delta+\Delta_\theta)^2 \,W^{\rm curv}_-W^{\rm curv}_+ \,+\, \nn\\
 && \q\q\q\q\q\q
  2\Delta(\Delta+\Delta_\theta) \left( V^\perp_-W^{\rm curv}_+\,+\, W^{\rm curv}_-V^\perp_+\right) \bigg) \,, \,\,\,\,
 \ea
 which comes with two different scalings with $R$, and gives rise to the contribution
  \ba
 S_{HJb}^{(2)}&=& \frac{1}{16\pi G} \frac{\Theta Y Z}{R \sqrt{1-\frac{\Theta^2}{4R^2}}} \sum_{k} (h'(k))^{\rm t} \circ \left( R^2 N^{(R^2)}_h+N^{(R^0)}_h\right) \circ h'(-k) .
 \ea
 \item  Transforming the first order in length variables of the Hamilton-Jacobi action (\ref{HJFO}) to metric variables  gives the following second order contribution in metric variables: 
 \ba
 S_{HJc}^{(2)}&=&\frac{YZ}{32\, G N_\theta}\!\! \sum_{k_\theta,k_y,k_z} \!\!\!\left( h'_{yy}(k) h'_{yy}(-k) + h'_{zz}(k) h'_{zz}(-k) \right) \nn\\
 &=&\frac{YZ}{32\, G} \left(    \frac{\Theta}{2\pi R}+  {\cal O}\left(\tfrac{1}{R^3}\right)          \right)\!\! \sum_{k_\theta,k_y,k_z} \!\!\!\left( h'_{yy}(k) h'_{yy}(-k) + h'_{zz}(k) h'_{zz}(-k) \right), \;\;
\ea
where we used the relation $\frac{\Theta^2}{2R^2}=1-\cos(\frac{2\pi}{N_\theta})$ to express $N_\theta$ as a function of $R$.

This term is  not invariant under boundary diffeomorphisms in the $y$- and $z$-directions. As mentioned above, this is due to having a non-vanishing first order for the Hamitlon-Jacobi action. To make these invariant under diffeomorphisms to second order requires compensating second order terms.

\item Another contribution is $S^{\partial \cal T}$ as computed in (\ref{BDTERM1}). Transformed to the rescaled boundary metric variables $h'$ we obtain  
\ba
S_{HJd}^{(2)}&=&S^{\partial \cal T}  \nn\\
&=& 
\scalemath{0.8}{ - \frac{ YZ}{64 G }\left(    \frac{\Theta}{2\pi R}+  {\cal O}\left(\tfrac{1}{R^3}\right)\! \right) \!\!\!  \sum_{k_\theta,k_y,k_z} 
\left( \begin{array}{c}  h'_{yy}(k)\\  h'_{zz}(k) \\  h'_{yz}(k) \end{array}\right)^{\rm t}\! \cdot\!
\left( \begin{array}{ccc}Y^2 \Delta_z  &  Yd_y^*+Z d_z-2 & 2 Y d_z \\ Yd_y+Zd_z^*-2 & Z^2 \Delta_y   & 2 Z d_y  \\ 2 Y d^*_z & 2 Z d^*_y & 8  \end{array} \right)  
\!\cdot\!
\left( \begin{array}{c}  h'_{yy}(-k)\\  h'_{zz}(-k) \\  h'_{yz}(-k) \end{array}\right)
} \nn\\
&\rightarrow&
\scalemath{0.8}{  \frac{ YZ}{ 32 G }  \frac{\Theta}{2\pi R}  \sum_{k_\theta,k_y,k_z} 
\left( \begin{array}{c}  h'_{yy}(k)\\  h'_{zz}(k) \\  h'_{yz}(k) \end{array}\right)^{\rm t} \cdot
\left( \begin{array}{ccc}0  &  1 & 0\\ 1 & 0  & 0 \\ 0 & 0 & -4  \end{array} \right)  
\cdot
\left( \begin{array}{c}  h'_{yy}(-k)\\  h'_{zz}(-k) \\  h'_{yz}(-k) \end{array}\right)\; . 
} 
\ea
The last line gives the terms that survive the continuum limit. As for the previous contribution $S_{HJc}^{(2)}$, this term is  not invariant under boundary diffeomorphisms in the $y$- and $z$-directions, nor is the sum of the two contributions $S_{HJc}^{(2)}+S_{HJd}^{(2)}$.

 \item The last contribution comes from the hyper-diagonal variables $h'_{\theta y z}$:
 \ba
  S_{HJe}^{(2)}&=&
   \frac{1}{16\pi G} \frac{\Theta Y Z}{R \sqrt{1-\frac{\Theta^2}{4R^2}}} \left( R^2 - \tfrac{1}{2}\Theta^2\right) \sum_{k}  h'_{\theta y z} h'_{\theta y z}  .
 \ea
 This seems to have a similar  scaling behavior to the first two contributions. However, we have defined $h'_{\theta y z}= \frac{1}{\Theta Y Z} h_{\theta y z}$, whereas for the standard metric variables we have $h'_{ab}= \frac{1}{\sqrt{H_{aa}H_{bb}}} h_{ab}$ with $h_{\theta y z}$ and $h_{ab}$  having the same scaling dependence on the (rescaled) length perturbations $\hat \ell_e=L_e \ell_e$.  Thus, in the continuum limit, where $\Theta=\epsilon\Theta_0, Y=\epsilon Y_0$, and $Z=\epsilon Z_0$, with $R$  constant, the $R^2$ term will a priori dominate over the other contributions.
 
 We will therefore assume that the length perturbations for the hyper-diagonal are chosen such that $h'_{\theta y z}=0$, and that  this contribution is vanishing. Indeed, the hyper-diagonal is a spurious variable, which does not have a representation in the continuum geometry, and as we have seen, can be decoupled from the remaining variables. We can thus interpret the condition that $h'_{\theta y z}$ falls off (sufficiently) fast, as a requirement on the discrete geometry such that it will allow for a sensible continuum limit. 
 \end{enumerate}

We can expand the action for large radii $R$ as  (assuming $|k_\theta|>>1$)
\ba
S_{HJ}^{(2)} &=& R \, (S_{HJ}^{(2)})^{(R^1)} + \frac{1}{R^{-1}} \, (S_{HJ}^{(2)})^{(R^{-1})} \,+\, {\cal O}(\tfrac{1}{R^3}) .
\ea
Assuming that the condition above on the hyper-diagonal fluctuations holds, the terms that dominate in the large $R$ limit come only from the $S_{HJa}^{(2)}$ and $S_{HJb}^{(2)}$ contributions:  
\ba
 (S_{HJ}^{(2)})^{(R^1)} &=&
  \frac{1}{16\pi G}    \frac{ \Theta Y Z}{4}\sum_{k} \bigg(
 - \frac{\Delta^2}{(\Delta-\Delta_\theta)} \left(v(k)v(-k) + \frac{1}{2}v(k)w(-k)+  \frac{1}{2}w(k)v(-k) \right) 
  + \nn\\
  &&\q\q\q\q\q\;\;
  \frac{1}{2} \Delta x(k) x(-k) -\frac{1}{4}  \frac{(\Delta+\Delta_\theta)^2}{(\Delta-\Delta_\theta)} w(k)w(-k)
  \bigg) \; ,
\ea
where we expanded the boundary metric fluctuation as $h'=vV^\perp+wW^{\rm curv} + xX^{\rm curv}\,+\, D^{\rm bdiff}$, with $D^{\rm bdiff}$ being an element in the subspace spanned by the diffeomorphisms tagent to the boundary. 

If we further restrict to boundary conditions that induce flat solutions, we obtain only one contribution, which results from integrating out the radial field:
\ba
{ (S_{HJ}^{(2)})^{(R^1)}}_{|{\rm flat} }&=&  - \frac{1}{16\pi G}   \frac{\Theta Y Z}{4}\sum_{k} 
  \frac{\Delta^2}{(\Delta_y+\Delta_z)} v(k)v(-k) .
\ea

Again, this part of the Hamilton-Jacobi function can be reproduced if we restrict to flat boundary conditions,  via a boundary field theory with action
\ba\label{bfflats1}
S_{\rm bf} =   \frac{R \Theta Y Z}{16\pi G}   \sum_{k}    
\left( \ell'_r(k) (\Delta_y+\Delta_z) \ell'_r(-k)  + \ell'_r(k) \, (b(k))^{\rm t} \circ h'(-k) + \ell'_r(-k)\, (h'(k))^{\rm t} \circ b(-k)  \right), \nn\\
\ea
where  
\ba
(b(k))^{\rm t} &=& -\frac{1}{4} (  2\left( \Delta_y+\Delta_z \right) \, ,\,\,  \left(2 \Delta_\theta+\Delta_z \right) \, ,\,\,   \left(2 \Delta_\theta+\Delta_y\right)  \, ,\, 
 2 d_{\theta } d_y \, ,\,\, 2 d_{\theta } d_z \, ,\,\,   d_y d_z \,\,)\nn\\
&=& -\frac{1}{4} \left(  2\Delta (V^\perp(k))^{\rm t} +  ( \Delta +\Delta_\theta)(W^{\text{curv}}(k))^{\rm t} \right) .
\ea

Note that
\ba
(a(k))^{\rm t} &:=&(   (\Delta_y +\Delta_z) \,,\,  (\Delta_\theta +\Delta_z)\,,\, (\Delta_\theta +\Delta_y)\, ,\,   d_\theta d_y \, , \,    d_\theta d_z \, ,\,   d_y d_z) \nn\\
&=& \Delta \left( (V^\perp(k))^{\rm t} \,+\,  (W^{\text{curv}}(k))^{\rm t} \right)
\ea
 agrees, on the flat sector, with a multiple of $b$
\ba
{b(k)}_{|\rm flat} \,=\, -\frac{1}{2}  {a(k)}_{|\rm flat} .
\ea
The boundary metric fluctuation $a(k)$ arises from a discretization\footnote{This discretization satisfies a consistency requirement,  namely that the vector $a$ is orthogonal to the diffeomorphisms tangential to the boundary, as one would expect from a quantity resulting from the discretization of a scalar density.} of the first order expansion of the Ricci scalar density $\sqrt{h}{\cal R}$, see appendix \ref{expansionRicci}.  

\section{The Boundary Action}\label{MainResult}

We thus come to the main result of our paper: restricted to the flat sector the discrete action (\ref{bfflats1}) coincides with a discretization of the second order expansion of the  following continuum action for a boundary field $\rho$
 \ba\label{contbdryf}
S_{\rm cbf} \,=\, \frac{R}{16\pi G} \int d^3x \sqrt{h} \left( -\rho ( h^{y y} \nabla_y \nabla_y + h^{z z} \nabla_z \nabla _z) \rho   -   {\cal R} \,\rho \right) .
\ea
 The expansion is around $\rho=0$ and around the flat boundary metric $h_{ab}=H_{ab}=\text{diag}(\Theta^2,Y^2,Z^2)$.

This effective action can be taken to describe the dynamics for flatly embeddable deformations of the boundary, that is, for  boundary gravitons. It is quite similar to an analogous action for three-dimensional gravity on a region with the topology $\text{Disk}\times  S^1$.  There one had also a degenerate kinematical term, as well as a coupling of the scalar field to the boundary Ricci-scalar.

Note that the degenerate kinematical term can be obtained by assuming a quadratic form $Q^{ab}=K^{ab} -h^{ab}K$ for the second derivatives. For the background four-metric $G_{\mu\nu}=(1,\tfrac{\Theta^2}{R^2}r^2,Y^2,Z^2)$  and the hyper-surface $r=R$, we have $Q^{ab} =-\frac{1}{R}\text{diag}(0,Y^{-2},Z^{-2})$, so that
\ba
Q^{ab}\nabla_a\nabla_b\,=\, -\frac{1}{R} ( h^{y y} \nabla_y \nabla_y + h^{z z} \nabla_z \nabla _z)   .
\ea
The $1/R$ factor can be absorbed by a rescaling of $\rho$ by $R$.  

\section{Hamilton-Jacobi Functional Contributions from the  $k_\theta=0$ and $k_\theta=\pm 1$ modes}\label{Lowmodes}

The contribution to the Hamilton-Jacobi action from the $k_\theta=0$ modes can be computed straightforwardly and yields
\ba
S_{HJ}^{(2)}(k_\theta=0)&=& \frac{1}{16\pi G} \frac{\Theta Y Z}{R \sqrt{1-\frac{\Theta^2}{4R^2}}} \sum_{k_y,k_z} \left((h'(k))^{\rm t} \circ  Q_h(k) \circ h'(-k)  + \ h'_{\theta yz}(k)M_{\theta y z}(k) h'_{\theta yz}(-k) \right)_{|k_\theta=0}\nn\\
\ea
with
\ba
Q_h(k)&=&  \frac{\Delta}{16} \Theta^2\left(\frac{1}{x} -1 \right) \left(     X^{\rm curv}_-X^{\rm curv}{(k)}_+ \,-\,   V^\perp_-W^{\rm curv}_+ \,-\,  W^{\rm curv}_- V^\perp_+\right) \nn \\
&&+  \frac{\Delta}{8}\Theta^2 \left( 1- \frac{1}{x}\right) V^\perp_-V^\perp_++ \frac{1}{2}\left(  X^{\rm curv}_- X^{\rm curv}_+ \,+\,  U^{\rm adiff}_- U^{\rm adiff}_+   \right) \nn \\
&& + \frac{1}{8} \Theta \sqrt{2(\Delta_y+\Delta_z)} \left( V^\perp_- U^{\rm adiff}{(k)}_+ + U^{\rm adiff}_- V^\perp_+ \right)   \nn \\
&& - \left(\frac{\Theta^2}{32x} \frac{(\Delta_y+\Delta_z)^2}{\left( \frac{8x}{\Theta^2} + \Delta_y + \Delta_z\right) } \right) W^{\rm curv}_- W^{\rm curv}_+  \,-\, 
  \frac{\Theta^2\Delta}{16x} \left(   V^\perp_- W^{\rm curv}_+\,+\,  W^{\rm curv}_- V^\perp_+\right)\,.\q\q\q
\ea
and 
\ba
M_{\theta y z}(k_\theta=0) =-\frac{1}{2} \left(\Theta^2- \frac{2\Theta^2}{2x} \right)  .\q\q
\ea

The dominant scaling in $R$  for $k_\theta=0$ coincides with that for general $k_\theta$, if we set $\Delta_\theta=0$ there. Thus the conclusions about the boundary field theory, which we defined in (\ref{bfflats1}), hold also in this case.

This situation appears, a priori, quite different if we consider the modes $k_\theta=\pm 1$. In this case we can use the general results found in section \ref{section:HJF}, if we set $\Delta_\theta=1/R^2$. As we discussed in section \ref{sec:low} the action for the radial field vanishes---the reason being that for $k_\theta=\pm 1$ the radial field can be taken as a gauge parameter. Indeed the pre-factor $(\Theta^2-2x/\Delta_\theta)$, which appears for $M_{rr}$ and $M_{rh}=B$, vanishes.
However, we have also a number of terms in $M_h^{(R^0)}$ with a $1/\Delta_\theta$ pre-factor, and these terms lead to an $R^2$ scaling in the $k_\theta=\pm 1$ case.

In fact, it turns out that the terms with a $1/\Delta_\theta$ coefficient coming from $N_h$ in (\ref{Nhdef}), i.e. those that come from integrating out the radial field, cancel all the terms with a $1/\Delta_\theta$ coefficient in $M_h$, that is, terms arising from integrating out all the other bulk fields:\footnote{The superscripts ${}^{(R^0)}$ and ${}^{(R^2)}$ refer only to the explicit $R$-dependence and do not take the implicit one via $\Delta_\theta$ into account.}
\ba
N^{(R^0)}_h&=&\frac{(\Delta_y+\Delta_z)}{16\Delta_\theta}\bigg( 4 V^\perp_- V^\perp_+ + W^{\rm curv}_- W^{\rm curv}_+ + 2\left(   V^\perp_- W^{\rm curv}_+ \,+\,  W^{\rm curv}_- V^\perp_+ \right) \bigg)
\,+\, {\cal O}((\Delta_\theta)^0) \nn\\
&=& - M^{(R^0)}_h  \,+\,  {\cal O}((\Delta_\theta)^0) .
\ea 
Thus, although the radial variables give a vanishing contribution for $k_\theta=\pm 1$---due to the cancellation between $N^{(R^0)}$ and $R^2 N^{(R^2)}$---this is compensated for by terms coming from $M^{(R^0)}_h$. This argument generalizes for the contributions from small $k_\theta \sim 1$ modes.

This shows that, even for $k_\theta=\pm 1$, the boundary field theory  (\ref{bfflats1}) will lead to the same Hamilton-Jacobi action---at leading order in the radius expansion and restricted to boundary fluctuations inducing flat solutions---as the gravitational bulk theory. We cannot, however, so easily identify the boundary field with the lengths of radial geodesics anymore.

\section{Null vectors of the Hessian matrix not related to gauge symmetries}\label{sec:singular}

For certain choices of $N_\theta$, $\gamma_y$, and $\gamma_z$, there will be momenta $(k_\theta,k_y,k_z)$ for which
\ba
\Delta_y+\Delta_z \,=\, \Delta-\Delta_\theta \,=\,  0 .
\ea
As we have  $\Delta_a\sim 2-2\cos( v_a/N_a)$ for $a\in \{y,z\}$, this happens when
\ba\label{ZerosInDelta}
v_y:=k_y-\frac{\Upsilon_y}{N_\theta}k_\theta =0 \,\, \text{mod}\,\, N_y  \q \text{and} \q v_z:= k_z-\frac{\Upsilon_z}{N_\theta}k_\theta =0 \,\, \text{mod}\,\, N_z  .
\ea
For $k_\theta\neq 0$, which we will assume here, a vanishing $(\Delta_y+\Delta_z)$ Laplacian  leads to null vectors for the bulk Hessian and thus zero's for its determinant. These null modes are not related to gauge symmetries. This is because these modes are only null for the bulk Hessian and not the full Hessian, which includes the boundary fluctuations. Indeed, when there are momenta for which $(\Delta_y+\Delta_z)=0$ there is only a solution to the linearized equations of motion if the boundary fluctuations satisfy
\ba\label{cond4}
(B(k))^{\rm t} \circ h'(-k) \,=\, 0   \q \text{and} \q (B(-k))^{\rm t} \circ h'(k) \,=\,0 .
\ea
Thus, if the twist angles are such that $\Delta_y+\Delta_z$ has a zero, we can find a solution to the linearized equation of motion only if (\ref{cond4}) is satisfied for all $k$ for which $\Delta_y+\Delta_z$ is zero. Note that it can also happen that a boundary metric that is in the image of the projector onto the flat sector might not allow for such a solution.

The same issue appears for the gravitational partition function for $(2+1)$-dimensional torus \cite{BonzomDittrich}. However, as discussed in \cite{qlh1}, there are indications that this is related to an artifact arising from the linearization, at least for finite radius $R$. 

The zeros of $(\Delta_y+\Delta_z)$ will lead to zero's for the determinant of the bulk Hessian, and thus singularities for the one-loop correction.\footnote{More precisely, the conditions required for the saddle-point approximation are not satisfied.} In \cite{BarnichOneLoop} these singularities for the $(2+1)$-dimensional partition function are dealt with by adding a small imaginary contribution to the (in that case single)  twist angle $\gamma=2\pi \Upsilon/N_\theta$. 

Let us find solutions for the equations (\ref{ZerosInDelta}) with $k_\theta \neq 0$. We will assume that $\Upsilon_y\neq 0$ and $\Upsilon_z \neq 0$.\footnote{For $\Upsilon_y=\Upsilon_z=0$ we will have a null mode $(k_y,k_z)=(0,0)$. The case where only one of the twisting angles vanishes can be easily deduced from the more general discussion.}  A necessary condition for finding such solutions to $v_y=0$ is that $p_y:=\text{GCD}(\Upsilon_y,N_\theta)>1$. Likewise a necessary condition for $v_z=0$ is  $p_z:=\text{GCD}(\Upsilon_z,N_\theta)>1$.  The values for $k_\theta$ for which $v_y$ vanishes, are given by
\ba
k_\theta\,=\, t_y \cdot q_y^N, 
\ea
where $t_y=1,2,\ldots, p_y-1$ and $q_y^N$ is defined by $N_\theta=q_y^N\cdot p_y$.  The associated solutions for $k_y$ are given by $k_y=t_y\cdot q_y^\Upsilon \, \text{mod}\, N_y$, where $\Upsilon_y=q_y^\Upsilon\cdot p_y$. Likewise we need for $v_z=0$
\ba
k_\theta\,=\, t_z \cdot q_z^N,
\ea
with $t_z=1,2,\ldots, p_z-1$ and $k_z=t_z\cdot q_z^\Upsilon  \, \text{mod}\, N_z$. Thus, to have $v_y=0$ and $v_z=0$, we need to satisfy for $t_y=1,2,\ldots, p_y-1$ and $t_z=1,2,\ldots,p_z-1$ the equation
\ba
t_y \cdot q_y^N \,=\,  t_z \cdot q_z^N .
\ea 

Note that for arbitrary twist angles $(\Upsilon_y,\Upsilon_z)\neq (0,0)$ we can always find discretizations for which no such zeros in the determinant of the bulk Hessian arise. To this end we just need to choose $N_\theta$ such that either $\text{GCD}(\Upsilon_y,N_\theta)=1$ or $\text{GCD}(\Upsilon_z,N_\theta)=1$.

We can also give a geometrical description of, for example, the condition $\text{GCD}(\Upsilon_y,N_\theta)=1$: on the hyper-torus, we consider a geodesic which starts at $(s_\theta,s_y,s_z)=(0,0,0)$ and for which (initially) $s_\theta$ and $s_z$ are constant. Going from the $s_y=N_y-1$ to the $s_y=N_y\equiv 0$ vertex, we have, however, to shift to the $s_\theta=\Upsilon_y$  vertex. If $\text{GCD}(\Upsilon_y,N_\theta)=1$ we only need one such geodesic to visit all vertices in the surface defined by $s_z=0$.  In the continuum theory the analogous condition is whether a geodesic, which goes initially along constant $(\theta=0)$- and $(z=0)$-coordinates, densely fills the torus defined by $(z=0)$.

\section{One-loop correction}  \label{sec:oneloop}

To find the one-loop correction we have to find the determinant of the Hessian describing the quadratic form in the bulk perturbations, which we will do in the following subsections.

The one loop correction is   given by  the Gaussian integral
\ba\label{PIDef}
{\cal C}_{1} \,=\,\int  \hat\mu_N(L)  \prod_{e \in \text{blk}} d\hat\ell_e  \,  \,  \exp\left( - S^{(2)}_{\rm blk}(\hat \ell) \right),
\ea
where $\hat\mu(L)$ is a measure factor for the $\hat \ell$ variables
\ba
\hat \mu_N(L) \,=\,     \mu_N(L) \prod_{e \in \text{blk}}   L_e 
\ea
and $\mu(L)$, the measure factor defined for the $\ell$ variables,  is discussed in section \ref{Section:Regge}. The bulk action is 
\ba
S^{(2)}_{\rm blk} \,=\, \frac{1}{\cal G} \sum_{k}  (\vec { \hat \ell}(k))^{\rm t} \cdot M_{\rm blk}(k) \cdot  \vec {\hat \ell}(k) (-k),
\ea
 the matrix $M_{\rm blk}(k)$ is detailed in appendix \ref{bulkaction}. Here ${\cal G}$ is shorthand for
\ba
{\cal G}\,=\,   16 \pi G \times 24 V_{\sigma}\,=\,  16 \pi G  R \Theta Y Z  \sqrt{1 -\frac{\Theta^2}{4R^2}}   .
\ea

There are seven types of bulk variables: 
\ba
(\vec{\hat{\ell}} )^{\rm t} \, =\, ( \hat{\ell}_r, \hat{\ell}_{ry}, \hat{\ell}_{rz},\hat{\ell}_{ryz},\hat{\ell}_{\varphi},\hat{\ell}_{\zeta},\hat{\ell}_{\varphi\zeta} )  .
\ea
The last three types, $( \hat{\ell}_{\varphi},\hat{\ell}_{\zeta},\hat{\ell}_{\varphi\zeta})$, describe edge lengths on the two-dimensional central axis and are only defined for $k_\theta=0$.

It is convenient to integrate out the various variables in steps, which we will describe in the following. We also have  to take care of the gauge symmetries arising for the $k_\theta=0$ and $k_\theta = \pm 1$ modes, this is best done separately.

\subsection{Contribution from $k_\theta=0$ }

The matrix $M_{\rm blk}(k_\theta=0)$ has two null eigenvectors (per $(k_y,k_z)$), which correspond to the two vertex translations  in the $y$- and $z$-direction of the vertices on the central two-dimensional axis. We described these null vectors in (\ref{bulkdiff1}).  From amongst the seven types of bulk variables we therefore integrate out only five, namely $\ell_r(0,k_y,k_z), \ell_{ry}(0,k_y,k_z),\ell_{rz}(0,k_y,k_z),\ell_{r y z}(0,k_y,k_z)$, and $\ell_{\varphi \zeta}(k_\psi,k_z)$.  The resulting effective action (if we allow for non-vanishing boundary fluctuations) does not depend on the remaining two variables $\ell_{\varphi }$ and $\ell_{\zeta}$. Below we will consider the measure over the gauge orbits resulting from the vertex translation symmetry, which will absorb the Lebesgue measure over the remaining variables $\ell_\varphi$ and $\ell_\zeta$.

Another peculiarity that appears for $k_\theta=0$ is that $M_{\rm blk}(k_\theta=0)$ has  one negative eigenvalue. This means that this contribution to the action is not bounded from below. This is a shadow of the well known conformal factor problem in general relativity.\footnote{One finds also a negative eigenvalue for the $k_\theta=0$ contribution in the 3D case, see \cite{BonzomDittrich}.} As usual, we formally rotate this eigenvalue to a positive sign.

It is not straightforward to isolate the eigenvector with the negative eigenvalue in the full matrix $M_{\rm blk}(k_\theta=0)$. However, one can first integrate out $\hat \ell_{\varphi \zeta}$ and then $\hat\ell_{r y z}$, which give a contribution of 
\ba
\text{det}_{\rm part 1}(k_y,k_z) \,=\, \frac{\Theta^4}{4} N_\theta
\ea
to the determinant of the Hessian $M_{\rm blk}(k_\theta=0)$. The remaining $3\times 3$ matrix has the following $2 \times 2$ block for the $(\hat\ell_{ry},\hat\ell_{rz})$ variables:
\ba
-\frac{\Theta^2 YZ}{2} \left(
\begin{array}{cc}
0& d^*_y d_z \\
d_y d^*_z&0
\end{array}
\right)
\ea
that has eigenvalues $\pm \frac{\Theta^2 YZ }{2} \sqrt{ \Delta_y \Delta_z}$.

Integrating out the $(\hat\ell_{ry},\hat\ell_{rz})$ variables, 
 we obtain the following $rr$-component for the matrix describing the effective action
\ba
(\tilde M_{\rm blk})_{rr}(k_\theta=0) \,\,=\,\, 2x Y^2 Z^2\left(\Delta_y+\Delta_z+ \frac{8x}{\Theta^2 }\right)  .
\ea
Already at this stage the effective action does not depend on $\hat \ell_\varphi$ or on $\hat \ell_\zeta$ anymore.

In summary, the product over the non-vanishing eigenvalues of $M_{\rm blk}(k_\theta=0)$ is given by
\ba
-\frac{N_\theta}{8} x \,\Theta^8 Y^4Z^4 \Delta_y \Delta_z   \left(\Delta_y+\Delta_z+ \frac{8x}{\Theta^2}\right)  .
\ea

We have also to consider that with our definition of the Fourier transform for $\ell_{\varphi\zeta}$ in (\ref{axisFT}) 
\ba
\prod_{s_y, s_z} d \hat \ell_{\varphi\zeta}(s_y,s_z) \,=\, \prod_{k_y ,k_z} \sqrt{N_\theta} \,d \hat \ell_{\varphi\zeta}(k_y,k_z). 
\ea
The contribution from integrating out the five types of variables $\hat\ell_r(0,k_y,k_z), \hat\ell_{ry}(0,k_y,k_z)$, $\hat \ell_{rz}(0,k_y,k_z),\hat \ell_{r y z}(0,k_y,k_z)$ and $\hat \ell_{\varphi\zeta}(k_y,k_z)$ is then given by
\ba
{\cal D}_0\,=\,\prod_{k_y,k_z} (2\pi {\cal G} )^{5/2}    \frac{2^{3/2}}{ x^{1/2}\Theta^4 Y^2 Z^2 \left( \Delta_y \Delta_z (\Delta_y + \Delta_z +\frac{8x}{\Theta^2})\right)^{1/2}} .
\ea

\subsection{Contribution from $k_\theta=\pm 1$ }

For $k_\theta=+1$ and for $k_\theta=-1$ we have $\Delta_\theta=2x/\Theta^2$. Using this relation one finds that the two matrices $M_{\rm blk}(k_\theta=+ 1)$ and $M_{\rm blk}(k_\theta=- 1)$ each have  one null eigenvector. These eigenvectors corresponds to the vertex translation symmetry for the vertices of the two-dimensional central axis, in the $(r,\theta)$ plane, see (\ref{bulkdiff2}). 

Hence, from amongst the four variables $( \hat{\ell}_r, \hat{\ell}_{ry}, \hat{\ell}_{rz},\hat{\ell}_{ryz} )$ we need only integrate out three and choose $( \hat{\ell}_{ry}, \hat{\ell}_{rz},\hat{\ell}_{ryz} )$. The determinant of the corresponding sub-matrix is given by 
\ba
\frac{1}{2} x\, \Theta^4 Y^2 Z^2  \left(\Delta_y+\Delta_z+ \frac{8x}{\Theta^2}\right) .
\ea

This leads to the following contribution to the one-loop correction
\ba
{\cal D}_{\pm 1}\,=\,\prod_{k_y,k_z} (2\pi {\cal G})^3 \frac{2}{x\, \Theta^4 Y^2 Z^2  \left(\Delta_y+\Delta_z+ \frac{8x}{\Theta^2}\right)   } .
\ea

\subsection{The measure over the gauge orbits}

We have not integrated over the variables $(\hat l_{\varphi}, \hat l_{\zeta})$  or $(\hat \ell_r(k_\theta=+1),\hat \ell_r(k_\theta=-1))$. But, after having performed the integrations outlined above, it is clear that the resulting action will be independent of these variables. Indeed these variables can be identified with gauge parameters for the vertex translation symmetries. In what follows, we identify a measure over the gauge orbits that will absorb the measure over these variables.

The gauge symmetry affects the vertices lying on the two-dimensional central axis. For each vertex $(s_y,s_z)$ we define the measure over the associated gauge orbit as
\ba
 \frac{1}{( 8\pi G)^2} \prod_{a=1,2,3,4}  d x^a(s_y,s_z),
\ea
where the $x^a(s_y,s_z)$ are Cartesian coordinates that describe the embedding of the given vertex into the flat solution. 
We identify $x^3$ and $x^4$ with $y$ and $z$ and have, to first order in the perturbations,  
\ba
\hat \ell_\varphi (s_y,s_z) &=&  Y( \delta x^3(s_y+1,s_z)- \delta x^3(s_y,s_z)) \ \  \Rightarrow    \ \ \hat \ell_\varphi (k_y,k_z)\,=\,  -Y^2 d_y \, \delta x^3(k_y,k_z), \nn\\
\hat \ell_\zeta (s_y,s_z) &=&  Z( \delta x^4(s_y,s_z+1)-\delta x^4(s_y,s_z)) \ \  \Rightarrow \ \     \hat \ell_\zeta (k_y,k_z)\,=\,  -Z^2 d_z  \,  \delta x^4(k_y,k_z)  . \ \ 
\ea
This gives for the measure\footnote{We remind the reader that due to our convention (\ref{axisFT}) we have $\prod_{s_y,s_z} d\hat \ell_\varphi(s_y,s_z) d\hat \ell_\zeta(s_y,s_z)=\prod_{k_y,k_z} N_\theta d\hat \ell_\varphi(k_y,k_z) d\hat \ell_\zeta(k_y,k_z)$ and $\prod_{s_y,s_z} d x^3 (s_y,s_z) d x^4(s_y,s_z)=\prod_{k_y,k_z} N_\theta d x^3(k_y,k_z) d x^4 (k_y,k_z)$.} over $x^3$ and $x^4$:  
\ba
\prod_{s_y,s_z} dx^3(s_y,s_z) dx^4(s_y,s_z)  &=& \prod_{k_y,k_z}  \frac{N_\theta}{Y^2Z^2\sqrt{\Delta_y \Delta_z}}  d\hat \ell_\varphi(k_y,k_z)  d\hat \ell_\zeta(k_y,k_z)  .
\ea

To discuss the vertex displacements in the $(r,\theta)$-plane we choose the $x^1$-axis parallel to the edges with the $\ell_r(s_\theta=0)$ variables. A displacement of a vertex at $(s_y,s_z)$ then results in a change of the $\hat \ell_r$ variable according to
\be
 ( R + R^{-1} \hat \ell_r(s_\theta, s_\psi, s_z))^2 = \left( R\cos(2\pi s_\theta/N_\theta) - \delta x^1 \right)^2 + \left( R\sin(2\pi s_\theta/N_\theta) - \delta x^2 \right)^2  .
\ee
To linear order this gives
\be
R^{-1}\,\hat \ell_r  \,\simeq \, - \cos(2\pi s_\theta/ N_\theta)\delta x^1 \,-\, \sin(2\pi s_\theta/N_\theta) \delta x^2  ,
\ee
and after Fourier transformation
\ba
R^{-1} \left(
\begin{array}{c}
\hat \ell_r(k_\theta=+1,k_y,k_z)\\
\hat \ell_r(k_\theta=-1,k_y,k_z)
\end{array}
\right)
\simeq
\frac{\sqrt{N_\theta}}{2} \left(
\begin{array}{cc}
-1&+ \i \\
-1 & -\i 
\end{array}\right)
\left( \begin{array}{c}
\delta x^1(k_y,k_z) \\
\delta x^2 (k_y,k_z)
\end{array}
\right) .
\ea
Thus, we have for the measure 
\be
\prod_{s_y,s_z} dx^1(s_y,s_z) dx^2(s_y,s_z) = \prod_{k_y, k_z} \frac{2}{R^2} d\hat\ell_r(+1,k_y, k_z)\, d\hat\ell_r(-1,k_y, k_z) .
\ee
In summary, the measure  over the gauge orbits of the vertex translation symmetry is given by 
\ba
&&\prod_{s_y,s_z} \frac{1}{(8\pi G)^2} dx^1 dx^2 dx^3 dx^4 \nn\\
&\,=\,& \prod_{k_y, k_z} \frac{1}{(8\pi G)^2}  \frac{2 N_\theta}{ R^2 Y^2 Z^2 } \frac{1}{\sqrt{\Delta_y \Delta_z}} d\hat \ell_\varphi(k_y, k_z) d \hat\ell_\zeta(k_y, k_z) d\hat\ell_r(+1,k_y, k_z)\, d\hat\ell_r(-1,k_y, k_z) \,.\q\;\;
\ea
Because we have to remove this integration measure from the path integral, it leads to the following contribution to the one-loop correction
\ba
{\cal D}_{\rm G}\,=\,  \prod_{k_y, k_z} (8\pi G)^2 \frac{ R^2 Y^2 Z^2 (\Delta_y \Delta_z)^{1/2}}{2N_\theta} .
\ea

\subsection{Contributions from $|k_\theta|>2$}

All that remains is to consider the modes with $|k_\theta|\geq 2$. For these modes there are no null vectors for $M_{\rm blk}$, and so we have to integrate out all four types of bulk variables $( \hat{\ell}_r, \hat{\ell}_{ry}, \hat{\ell}_{rz},\hat{\ell}_{ryz} )$. For the latter three types of variables we obtain a determinant
\ba
\frac{1}{4}  \Theta^6 Y^2 Z^2  \, \Delta_\theta\, \left( 4\Delta_\theta + \Delta_y+ \Delta_z\right)  .
\ea
The  resulting effective action for the $\hat \ell_r$ variable is described by the coefficient
\ba
(\tilde M_{\rm blk})_{rr}\,=\, \frac{2x}{\Theta^2} Y^2 Z^2\left(\Theta^2-\frac{ 2x }{\Delta_\theta}\right)  \left( \Delta_y+\Delta_z\right)   , 
\ea
which agrees with (\ref{Mrr1}) after taking into account the different scalings of $\ell'_r$ and $\hat \ell_r$ and an additional overall factor $V_{\rm cube}^{2}=\Theta^2 Y^2 Z^2$.

Note that  the $(\Delta_y+\Delta_z)$ factor in $(\tilde M_{\rm blk})_{rr}$ might have zero's, which we described in section 
 \ref{sec:singular}. As these zero's are not related to a gauge symmetry, they lead to singularities for the one-loop correction. 

The contribution of the $|k_\theta|\geq 2$ modes to the one-loop correction is given by
\ba
{\cal D}_{\geq 2}&=& \prod_{k_{\theta}=2}^{N_\theta-2} \prod_{k_y,k_z} (2\pi {\cal G})^{2}\frac{2^{3/2}}{x^{1/2} \Theta^2 Y^2Z^2}\frac{1}{\Delta_\theta^{1/2} (\Theta^2-\frac{2x}{\Delta_\theta})^{1/2} (4\Delta_\theta+\Delta_y+\Delta_z)^{1/2}(\Delta_y+\Delta_z)^{1/2} } \, .\nn
\ea

\subsection{Final result for the one-loop correction}\label{section:oneloop}

To compute the product over $k_\theta$-modes  of $\Delta_\theta (\Theta^2-\tfrac{2x}{\Delta_\theta})$ we use the results
\ba
\prod_{k_\theta=1}^{N_\theta-1}\Theta^2\Delta_\theta \,=\, 2 , \q \text{and} \q \prod_{k_\theta=2}^{N_\theta-2}\left(1-\frac{2x}{\Theta^2 \Delta_\theta}\right) \,=\, \frac{1}{4-2x}  .
\ea

This leaves us with the following expression for the one-loop correction
\ba\label{resultc1}
{\cal C}_1&=&\mu_N(L) \left( \prod_{e\in \text{blk}} L_e\right) {\cal D}_0 {\cal D}_{\pm 1} {\cal D}_{\geq 2}{\cal D}_{\rm G}\nn\\
&=& {\cal N}  \left[\prod_{k_y,k_z} \frac{1}{\left( (\Delta_y+\Delta_z)_{|k_\theta=0}+\frac{4}{R^2}\right)^{1/2}} \right]
\left[ \prod_{k_\theta=1}^{N_\theta-1}\prod_{k_y,k_z}   \frac{1}{ \left( 4 \Delta_\theta+\Delta_y +\Delta_z\right)^{1/2} } \right]\times \nn\\
&&\;\; \;\;\left[ \prod_{k_\theta=2}^{N_\theta-2}\prod_{k_y,k_z}  \frac{1}{ \left( \Delta_y + \Delta_z\right)^{1/2} } \right] .
\ea
The factor ${\cal N}$ is given by
\ba
{\cal N} &=& \mu_N(L) \left( \prod_{e\in \text{blk}} L_e\right) (2\pi {\cal G})^{2N_yN_z(N_\theta-1/4)} (8\pi G)^{2N_yN_z} 2^{2 N_yN_z(N_\theta-\tfrac{3}{2})} N_\theta^{-2N_yN_z}  \times \nn\\ 
&&\q\q\q\q\left(4-\tfrac{\Theta^2}{R^2}\right)^{\tfrac{1}{2}N_yN_z} R^{N_yN_zN_\theta}\Theta^{-N_yN_zN_\theta} (YZ)^{-2N_yN_z(N_\theta-2)} ,
\ea
where $\mu_N(L)$ is a choice for the measure in the background path integral, see (\ref{PIDef}).\\[1mm]

The one-loop approximation for the path integral is given by
\ba
{\cal Z}_1\,=\, {\cal C}_1 \, \exp\left( - S^{[2]}_{\rm HJ}(L,\ell_{\rm bdry})\right) ,
\ea
where $S^{[2]}_{\rm HJ}(L,\ell_{\rm bdry})$ is the second order approximation of the Hamilton-Jacobi action. Setting the boundary fluctuations $\{\ell_{\rm bdry}\}$ to zero, we will have
\ba
{\cal Z}_1\,=\, {\cal C}_1 \, \exp\left( \frac{\alpha \beta}{4G} \right)  .
\ea
The classical, on-shell action does not depend on the twist angles $(\gamma_y,\gamma_z)$. The one-loop correction ${\cal C}_1$ does, however, depend on these twists, via their appearance in the Laplacians $\Delta_y$ and $\Delta_z$ respectively. In particular, the last factor in (\ref{resultc1}) will be singular if $(\Delta_y+\Delta_z)$ has zero's. As discussed in section \ref{sec:singular}, the appearance of such zero's depends on the twist angles, as well as on $N_\theta$.

Let us emphasize again that the one-loop correction, as we have calculated it here, depends on our choice of triangulation. The singularities we have found will, however,  persist if we consider finer discretizations. The reason for this is that these singularities result from the effective boundary field theory for the radial length variables. This effective field theory, if restricted to the flat sector (which, in particular, includes the boundary condition where we set all boundary fluctuations to zero), is invariant under changes of the bulk triangulation. That is, even if we would start with a much finer triangulation, we would find again, via a coarse graining procedure, the same boundary field theory for the radial length variables. Integrating out these radial length variables, we would encounter the same kind of zero's for the determinant of its Hessian, which, in turn, lead to singularities for the one-loop correction.

Indeed, the existence of the singularities can be traced back to the fact that one cannot find solutions to the linearized  Einstein's equations for certain boundary conditions. This feature  also exists for the continuum linearized Einstein's equations; the obstruction is topological in nature, as explained at the end of section \ref{sec:singular}.

\section{Discussion} \label{discussion}

In this work we derived a boundary theory that encodes the dynamics of boundary gravitons in 4D gravity.  These boundary gravitons  describe the deformations of the boundary under diffeomorphisms. This geometrical interpretation motivates our choice of boundary field, namely the geodesic distance from a given point on the boundary to a central axis.  

As background spacetime we have worked with a solid hyper-torus with radius $R$. When restricted to the flat sector of boundary metrics that lead to a flat 4D solution, and in the large radius limit, the action encoding the dynamics of the three-dimensional ($d=3$) boundary theory is given by the second order approximation to
\ba\label{Disc1}
S  \sim \int d^d x  \sqrt{h}  \left( \phi \, Q^{ab} \nabla_a \nabla_b  \phi - {\cal R} \phi  \right)  , 
\ea
where the kinematical term is characterized by the trace-reversed extrinsic curvature $Q^{ab}=K^{ab}-h^{ab}K$ and ${\cal R}$ is the Ricci scalar on the boundary.

The same boundary action (with $d=2$) was found for a solid torus in 3D gravity. In this case the flat sector includes all boundary metrics,\footnote{As described in section \ref{sec:singular}, in 4D there are certain boundary metrics that do not have a solution in the linearized theory.} and thus the boundary theory encodes all of the dynamics of 3D gravity.

The quadratic form $Q^{ab}$ appearing in the action (\ref{Disc1}) is, for the background spacetime we have considered here, degenerate. However, when a Dehn-twist is included in the hyper-torus it leads to a twist action on the leftover laplacian. For this reason the zero's that appear for the laplacian (as a function of the momenta) depend on the twist parameters. 

These zero's lead to singularities for the one-loop correction, which in the 3D case completely characterize its dependence on the twist angles. In 4D we are not yet able to compute the full one-loop correction, but can conclude that these singularities will also feature in the continuum partition function.   

Another interesting feature of the boundary theory results from the bulk diffeomorphism symmetry of gravity. These diffeomorphisms can change the position of the central axis or central point from which the distance to the boundary points is determined. Indeed, with our background spacetime this symmetry affects the lowest modes in the angular direction, $k_\theta=0$ and $k_\theta=\pm 1$, of the boundary field. We discussed the resulting subtleties for the boundary theory in section \ref{Lowmodes}.   In particular, the action for the boundary field vanishes for $k_\theta=\pm 1$, due to the fact that one understands the radial field for these modes as pure gauge.

Integrating out the boundary field from the boundary action (\ref{Disc1}) we obtain the Hamilton-Jacobi function, that is, the on-shell action, for 4D gravity restricted to the flat sector (i.e. those boundary metrics that induce a flat solution). We have also computed the on-shell action for general boundary metrics, albeit, due to the coarse triangulation we have used, in a severe truncation. The result is quite complicated, but simplifies drastically in the large radius limit. Future work will show whether this result persists when the bulk triangulation is refined. \\[1mm]
 
This brings us to a number of directions opened up by this work:\\[1mm]

As mentioned in the introduction, instead of 4D gravity, we can consider a (quantum) theory of 4D flat space, which to some extent is quite similar to 3D gravity.  Such a theory has been proposed  in \cite{BaratinFreidel4D} (see also \cite{BaratinFreidelq}).  Adopting a form that is more suited to our context, the partition function has the same kinematical ingredients as Regge calculus and can be written as
\ba\label{BFRD}
{\cal Z}(l_\text{bdry})\,=\,\int \mu_{\rm inv}(l)  \exp( \i S_R) \prod_{e \in {\cal T}^\circ} dl_e  \, \prod_{t \in {\cal T}^\circ} \delta(\epsilon_t ) .
\ea
Here $S_R$ is the Regge action, which due to the delta functions in (\ref{BFRD}), reduces to a boundary term. As we have restricted to flat solutions, the on-shell action will be invariant under changes of the bulk triangulation. The measure $\mu_{\rm inv}(l)$ can also be chosen such that the partition function is bulk triangulation invariant.  The delta-functions appearing in (\ref{BFRD}) might overlap and produce divergences, but these can be consistently removed in such a way that the partition function is triangulation invariant, see \cite{BaratinFreidel4D}.

Clearly (\ref{BFRD}) describes the embedding of the boundary hypersurface into flat space, with each boundary configuration that allows for a flat bulk solution, weighted by the (Regge) gravity boundary term.  In this sense, this theory is similar to 3D gravity, for which we discussed a similar interpretation in the introduction, see also \cite{BaratinFreidel3D}.

The partition function (\ref{BFRD}) will vanish outside the flat sector, in other words, for those boundary metrics that do not induce a flat solution. The on-shell action will, on the flat sector, coincide between this theory and 4D gravity and so the boundary theory we have identified for the flat sector of 4D gravity will also be a boundary theory for (\ref{BFRD}).

However, the one-loop correction will differ between the two theories. For (\ref{BFRD}) we can determine this one-loop correction using an arbitrarily coarse triangulation \cite{toappear1}. 

This theory of quantum flat space can be formulated as a Topological Quantum Field Theory (TQFT)  based on a two-category \cite{BaratinFreidelq}. TQFT's are proposed to play an essential role also for 4D quantum gravity, e.g. \cite{SmolinTQFT,BaerenzBarrett,DG16,QVacuum4D}, but most work is so far concentrated on BF-like TQFT's \cite{PerezSpinFoams,NewVacuum1,NewVacuum2}, which start from an enlarged space of generalized simplicial geometries \cite{AreaAngle,DittRyan1, Bonzom,TwistedGeom,DittRyan2}. A key problem is to devise either a mechanism to restrict back to proper geometric configurations \cite{PerezSpinFoams,BarrettCrane,Lev-Spez,EPRL,FK,HolomorphicSF,BOM} or to find a dynamical principle for these generalized simplicial geometries \cite{Simone}. It will be fruitful to explore alternatives, such as the one just discussed, even if these end up `only' describing flat space.
\\[1mm]

In this work we have considered a spacetime with the topology of a solid hyper-torus. We have found a particular form (\ref{Disc1})  for the boundary theory, which turns out to hold both for the 4D spacetime and for the 3D solid torus.  It will be interesting to know whether the same boundary theory holds also for more general topologies. In particular, it would be interesting to consider boundaries with topology $S^2\times S^1$, as this would include Euclidean black holes. This case would be relevant for studying connections to the BMS symmetry shown to exist for the 3D theory \cite{BarnichOneLoop,Oblak1,Oblak2}. Another generalization would be to add a cosmological constant. This can also be considered within Regge calculus, if one uses homogeneously curved building blocks \cite{NewRegge,Haggard:15,Haggard:16,Haggard:162}. 

Here we constructed the boundary theory as the effective theory of geodesic distances from the boundary to some central point(s). One could also look for other geometric variables that describe the embedding of the 3D boundary into the 4D solutions. In the 3D case one can find boundary theories based on different geometric variables \cite{qlh1,qlh2,qlh3,qlhA,WW2}. To this end, one uses versions of Regge calculus based on other sets of variables than the edge lengths, e.g. areas and angles \cite{BarrettFO,AsanteDittrichHaggard18,AreaAngle,NewRegge}. Another choice, possibly more suited for Lorentzian signature, would be variables related to spinors or twistors \cite{WW1,WW2}. 

For the 3D theory, the boundary field leads to a similar encoding of the bulk geometry as in the Ryu-Takayanagi proposal \cite{RyuTak}, which in 3D is based on geodesic distances between boundary points  (see also  \cite{Scullyetal}). In 4D, the Ryu-Takayanagi proposal would, however, involve the area of minimal surfaces, whereas here we are still using geodesic lengths. It could be interesting to derive a boundary theory based on the areas of minimal surfaces.
\\[1mm]

We have been focused on the flat sector of 4D gravity, which allowed us to work with an arbitrarily coarse bulk triangulation. However, this work also provides the setup to consider refinements of the bulk triangulation.  Using coarse graining methods applicable to Regge calculus \cite{Improve,BahrDittrichHe,Ditt12}, we can construct a renormalization flow for discrete gravity.  The model we used in this work has proven to lead to manageable computations. It seems, therefore, to allow for a further evaluation of the dynamics, which with increasing bulk refinement, will also include more and more curvature degrees of freedom. The unusual feature of the setup here is that one can take already in advance the continuum limit for the boundary. This might actually simplify the study of the coarse graining flow, as it allows us to identify continuum geometric quantities and to consider the flow of an action that is a functional of these geometric quantities. This might allow one to identify relevant and irrelevant geometric variables, which would also be useful for coarse graining other theories, e.g. spin foam models \cite{HolSF,DittrichReview14,BahrR,DecTNW,DelcampDittrich,BSGiov}.

The coarse graining flow can also help to identify a measure for 4D Regge calculus that is invariant under bulk triangulation changes. As shown in \cite{ReggeMeasure2},  such a measure must be non-local and is difficult to guess. It will be interesting to see whether such a coarse graining flow, which only affects refinement in the radial directions, also leads to such a non-local (fixed point) measure.  Alternatively, one can restrict to a local form of the measure and attempt to find the best local approximation to an invariant measure \cite{ReggeMeasure2,BahrSteinhaus1,BahrSteinhaus2}.\\[1mm]

In summary, we have identified a sector of 4D gravity---the flat sector---for which we can (more) easily access the dynamics. Although it describes a spacetime without (bulk) graviton excitations, this sector has as rich a dynamics as 3D gravity. In particular, it describes how the boundary is embedded into flat spacetime. We have identified a theory, defined on the boundary itself, that encodes this dynamics, and found astonishing parallels between the 3D and 4D case. 

The central aim of this work was to find the (one-loop) partition function for non-asymptotic  `generalized'  boundaries \cite{Oeckl}. Such partition functions can serve as (semiclassical) vacuum functionals. Understanding the vacuum functionals for such generalized boundaries  will also be crucial for coarse graining and renormalization in quantum gravity  \cite{TimeEvol,DittrichReview14}.  We hope that this will be the starting point for a more systematic understanding of the semiclassical vacuum functional for generalized boundaries in quantum gravity.

\section*{Acknowledgements}

 HMH thanks the Perimeter Institute for Theoretical Physics for generous sabbatical support. SKA is supported by an NSERC grant awarded to BD. This work is  supported  by  Perimeter  Institute  for  Theoretical  Physics.   Research  at  Perimeter  Institute is supported by the Government of Canada through Industry Canada and by the Province of Ontario through the Ministry of Research and Innovation.

\appendix

\section{Transformation from lengths to metric variables}\label{app:lengthmetric}

We want to interpret the boundary theory as a theory coupled to the boundary metric. We therefore need to change the (boundary) edge length fluctuation variables $\ell_e$ into metric fluctuation variables $h_{ab} \equiv g_{ab} - H_{ab} $, with $H_{ab}$ the background (boundary) metric.   Fixing a set of edge vectors $(e_\theta, e_y, e_z)$ that describe a discretization cell, the boundary length  variables $\ell_b$ are related to the metric variables $ h_{ab}$ as follows: 
\begin{align*}
&(H_{ab} + h_{ab})e^a_\theta e^b_\theta = \left( \Theta+ \ell_\theta \right)^2, \q (H_{ab} + h_{ab})(e^a_\theta +e^a_y)(e^b_\theta + e^b_y) = \left( \sqrt{ \Theta^2+Y^2 }+ \ell_{\theta y} \right)^2 ,    \\
&(H_{ab} + h_{ab})e^a_y e^b_y = \left( Y+ \ell_y \right)^2, \q (H_{ab} + h_{ab})(e^a_\theta +e^a_z)(e^b_\theta + e^b_y) = \left( \sqrt{ \Theta^2+Z^2 }+ \ell_{\theta z} \right)^2 , \\
&(H_{ab} + h_{ab})e^a_z e^b_z = \left( Z+ \ell_z \right)^2, \q (H_{ab} + h_{ab})(e^a_y +e^a_z)(e^b_\theta + e^b_y) = \left( \sqrt{ Y^2+Z^2 }+ \ell_{y z} \right)^2, \\
&\text{and} \q (H_{ab} + h_{ab})(e^a_\theta +e^a_y+e^a_z)(e^b_\theta +e^b_y+e^b_z) = \left( \sqrt{ \Theta^2+Y^2+ Z^2 }+ \ell_{\theta y z} \right)^2  .
\end{align*}
The background boundary metric (with respect to the basis vectors $(e_\theta, e_y, e_z)$) is given by $H_{ab} = \text{diag}(\Theta^2, Y^2,Z^2)$. These relations express the six metric components in terms of the seven length variables per vertex, and hence one of the length variables is redundant. Following \cite{Williams3D,DittFreidSpez}, we introduce an auxilliary `metric variable' $h_{\theta y z}$ and have seven discrete metric variables $h_e \in \{ h_{ab} , h_{\theta y z}\}$ with $a,b \in \{ \theta, y ,z \}$. There is a transformation such that the variable $h_{\theta y z}$ decouples from the boundary effective action. In the discrete Fourier transformed picture, the transformation between the discrete metric variables $h_e$ and the rescaled length variables $\hat{\ell_e} = L_e \ell_e$ that decouples $h_{\theta y z}$ is given by 
\be
h_{e}(k) = \sum_{e} T_{ee'} (k) \,\hat{\ell}_{e'} (k)+ \mathcal{O} (\hat{\ell}_e{}^2),
\ee
with
\be
T_{ee'}(k)=  \begin{pmatrix} 2 & 0 & 0 & 0 & 0 &0 &0  \\ 0 & 2 & 0 & 0 & 0 &0 &0  \\ 0 & 0 & 2 & 0 & 0 &0 &0  \\ -1 & -1 & 0 & 1 & 0 &0 &0  \\ -1 & 0 & -1 & 0 & 1 &0 &0  \\ 0 & -1 & -1 & 0 & 0 & 1 &0 \\ \frac{1}{2}(\omega_y + \omega_z) & \frac{1}{2}(\omega_\theta + \omega_z) & \frac{1}{2}(\omega_\theta + \omega_y) & -\frac{1}{2}(1 + \omega_z) & -\frac{1}{2}(1 + \omega_y) &- \frac{1}{2}(1 + \omega_\theta) & 1 \end{pmatrix}, 
\ee
where the $\omega_a$'s  are the discrete Fourier coefficients defined in \eqref{omegas}.

\section{Expansion of the Ricci scalar}\label{expansionRicci}

Here we compute the first order expansion of the densitized Ricci scalar  $\sqrt{h}{\cal R}$ around the (flat) background boundary metric.

As is well known the first order variation of the densitized Ricci scalar is given by
\ba
\delta(\sqrt{h}{\cal R})&=&\sqrt{H}\left(\tfrac{1}{2}H^{ab} \,{}^{B}\!{\cal R} - \,{}^{B}\!{\cal R}^{ab}\right)\delta h_{ab} + \sqrt{H} \nabla^a\left( \nabla^b \delta h_{ab}- H^{bc} \nabla_a \delta h_{bc} \right) ,
\ea
where ${}^{B}\!{\cal R}_{ab}$ and ${}^{B}\!{\cal R}$ denote the Ricci tensor and Ricci scalar of the background metric $H_{ab}$.

We are considering a flat background $H_{ab}=\text{diag}(\Theta^2,Y^2,Z^2)$, and thus $\nabla_a=\partial_a$, and ${}^{B}\!{\cal R}_{ab}=0$. Hence we have for our background
\ba
\delta(\sqrt{h}{\cal R})&=& \sqrt{H} \left(H^{ac}H^{bd} - H^{ab}H^{cd} \right) \partial_{a} \partial_b  \delta h_{cd} .
\ea
Introducing the scaled derivatives $\partial'_a=\sqrt{H^{aa}}\partial_a$ and variables $\delta h'_{ab}=\sqrt{H^{aa}}\sqrt{H^{bb}} \delta h_{ab}$ this gives, in component form, 
\ba
\delta(\sqrt{h}{\cal R})&=&\,\, \Theta Y\! Z \left( -(\partial'^2_y +\partial'^2_z) \delta h'_{\theta\theta}  -(\partial'^2_\theta +\partial'^2_z) \delta h'_{yy} -(\partial'^2_\theta +\partial'^2_y) \delta h'_{zz} \right)  \nn\\
&&+ 2 \Theta Y\! Z \,   \left( \partial'_\theta \partial'_y \delta h'_{\theta y} +\partial'_\theta \partial'_z \delta h'_{\theta z} +\partial'_y \partial'_z \delta h'_{yz} \right) .
\ea

\section{The bulk action}\label{bulkaction}

To compute the second order of the Regge action for length perturbations in general dimensions $D$, we will need the derivatives of the volumes $V_h$ of  the $(D-2)$-simplices (the hinges)  and  the derivatives of the dihedral angles $\theta_h$ at these hinges. Given a $d$-simplex $\sigma$ labelled with vertices $\{ 0,1,\cdots,d \}$, its volume as a function of its edge lengths is given in terms of the Cayley-Menger determinant 
\be
V_\sigma^2 = \frac{(-1)^{d-1}}{2^d d!} \cdot \det \begin{pmatrix} 0 &1 & 1 & 1& \hdots &1 \\ 1&0 & l_{01}^2 & l_{02} ^2& \hdots & l_{0d}^2  \\ 1& l_{01}^2 & 0  & l_{12}^2 & \hdots & l_{1d}^2 \\ 1& l_{02}^2 & l_{12}^2 & 0& \hdots & l_{2d}^2 \\ \vdots & \vdots & \vdots  & \vdots& \ddots & \vdots  \\ 1& l_{0d}^2  & l_{1d}^2 & l_{2d}^2 & \hdots & 0 \end{pmatrix},
\ee
where $l_{ij}$ is the edge length between vertices $i$ and $j$. From this formula one can determine easily the derivatives of the simplex volume with respect to the length variables.

A general formula for the derivatives of the dihedral angles in a simplex $\sigma$ is given by \cite{DittFreidSpez}
\be
\frac{\partial \hat{\theta}_{ij}}{\partial l_{kl}} = \frac{1}{d^2}  \frac{\hat{V}_k \hat{V}_l}{ V_\sigma^2} \frac{l_{kl}}{\sin \hat{\theta}_{ij}} \left( \cos \hat{\theta}_{ik} \cos \hat{\theta}_{jl} + \cos \hat{\theta}_{il} \cos \hat{\theta}_{jk} + \cos \hat{\theta}_{ij} ( \cos \hat{\theta}\cos \hat{\theta}_{ik} \cos \hat{\theta}_{il} + \cos \hat{\theta}_{jk} \cos \hat{\theta}_{jl} ) \right)
\ee
where $\hat{V}_{k}$ is the volume of the $(D-1)$-simplex, which is obtained by dropping the vertex $k$ in the simplex $\sigma$ and $\hat{\theta}_{ij}$ is the dihedral angle between the two faces of the simplex that are opposite  the vertices $i$ and $j$. 

Using these formulas, we can compute, for our choice of triangulation $\cal T$ of the solid hyper-torus, the Hessian matrix
\be \label{hess-mat}
H^{\cal T}_{ee'}=  \sum_{\sigma} \left.\left( \sum_{t\subset \sigma} \frac{\partial A_t}{\partial l_e} \, \frac{\partial \theta^\sigma_t}{\partial l_{e'}}  \right) \right|_{l_e = L_e}  = \frac{L_e L_{e'}}{24 V_\sigma} M_{ee'},
\ee
which appears in the second order Regge action. To this end we employ the discrete Fourier transform that diagonalizes the Hessian into blocks labelled by momenta $k = (k_\theta, k_y,k_z)$. We only give here the bulk part of the Hessian, which is 
\ba\label{Mbulk}
M_{\rm blk} (k)&=&\scalemath{0.8}{ \left(
\begin{array}{ccc}
 \Theta ^2+\Theta ^2\left(Y^2+Z^2\right) \Delta _{\theta } & \cdots &  \cdots  \\
 Z^2 \left( 2xY d^*_y - \Theta^2  \Delta_{\theta}\right) - \frac{\Theta^2}{2}\left(1 + \omega_y^{-1} \right) &
  \Theta ^2 +  \Theta ^2 Z^2 \Delta _{\theta } & \cdots  \\
  Y^2 \left( 2xZ d^*_z - \Theta^2  \Delta_{\theta}\right) - \frac{\Theta^2}{2}\left(1 + \omega_z^{-1} \right) &
   \frac{1}{2} \Theta ^2 \left(\omega _y+\omega _z^{-1}\right) &   \Theta ^2 +  \Theta ^2 Y^2 \Delta _{\theta }  \\
 \frac{\Theta ^2}{2} \left(\omega_y^{-1}+\omega _z^{-1} \right) & - \frac{\Theta ^2}{2} \left(1+\omega _z^{-1} \right) & - \frac{\Theta ^2}{2} \left(1+\omega _y^{-1} \right)   \\
 - \delta _{0,k _{\theta }}\sqrt{N_{\theta }}  \left( x Z^2- \frac{\Theta ^2}{4} \omega _z\right) \left( 1+ \omega
   _{\theta }^{-1}\right)& - \delta _{0,k _{\theta }}\sqrt{N_{\theta }} \, \frac{\Theta ^2 \left(1+\omega _{\theta }^{-1}\right)}{4}   & - \delta _{0,k _{\theta }}\sqrt{N_{\theta }} \, \frac{\Theta ^2 \left(1+\omega _{\theta }^{-1}\right)\omega_z}{4}   \\
 - \delta _{0,k _{\theta }}\sqrt{N_{\theta }}  \left( x Y^2- \frac{\Theta ^2}{4} \omega _y\right) \left( 1+ \omega
   _{\theta }^{-1}\right)   & -\delta _{0,k _{\theta }}\sqrt{N_{\theta }} \, \frac{\Theta ^2 \left(1+\omega _{\theta }^{-1}\right)\omega_y}{4}  &\delta _{0,k _{\theta }}\sqrt{N_{\theta }} \, \frac{\Theta ^2 \left(1+\omega _{\theta }^{-1}\right)}{4}  \\
\delta _{0,k _{\theta }}\sqrt{N_{\theta }} \, \frac{\Theta ^2 \left(1+\omega _{\theta }^{-1}\right)}{4}  & \delta _{0,k _{\theta }}\sqrt{N_{\theta }} \, \frac{\Theta ^2 \left(1+\omega _{\theta }^{-1}\right)}{4}  & -\delta _{0,k _{\theta }}\sqrt{N_{\theta }} \, \frac{\Theta ^2 \left(1+\omega _{\theta }^{-1}\right)}{4}  \\
\end{array}
\right.} \q \q \ \  \nn \\
&& \scalemath{0.8}{\left. \q \q \q \q 
\begin{array}{cccc}
 \cdots & \cdots & \cdots & \cdots \\
\cdots &  \cdots & \cdots & \cdots \\
\cdots & \cdots & \cdots & \cdots \\
\Theta ^2 & \cdots & \cdots & \cdots \\
\delta _{0,k _{\theta }}\sqrt{N_{\theta }} \, \frac{\Theta ^2 \left(1+\omega _{\theta }^{-1}\right)}{4} &  \frac{1}{2} \Theta ^2 \left( \delta _{0,k _{\theta }} \right)^2 N_{\theta }& \cdots & \cdots \\
-\delta _{0,k _{\theta }}\sqrt{N_{\theta }} \, \frac{\Theta ^2 \left(1+\omega _{\theta }^{-1}\right)}{4}  & 0   & \frac{1}{2} \Theta ^2 \left( \delta _{0,k _{\theta }} \right)^2 N_{\theta }  & \cdots \\
 -\delta _{0,k _{\theta }}\sqrt{N_{\theta }} \, \frac{\Theta ^2 \left(1+\omega _{\theta }^{-1}\right)}{4}  & -\frac{1}{2} \Theta ^2 \left( \delta _{0,k _{\theta }} \right)^2 N_{\theta } & -\frac{1}{2} \Theta ^2 \left( \delta _{0,k _{\theta }} \right)^2 N_{\theta} & \frac{1}{2} \Theta ^2 \left( \delta _{0,k _{\theta }} \right)^2 N_{\theta } \\
\end{array}
\right) }.
\ea
Here the variables have the ordering $(\vec{\hat{\ell}} )^{\rm t} \, =\, ( \hat{\ell}_r, \hat{\ell}_{ry}, \hat{\ell}_{rz},\hat{\ell}_{ryz},\hat{\ell}_{\varphi},\hat{\ell}_{\zeta},\hat{\ell}_{\varphi\zeta} ) $.
The definitions of the various phases $\omega_\theta,$ etc., and difference operators $\Delta_\theta,d_y,$ etc., can be found in Eqs.  \eqref{d-def}. We use the abbreviation $x = \frac{\Theta^2}{2R^2}$. The missing entries of the matrix can be found by imposing hermiticity.

\end{document}